# Estimating the Triaxiality of Massive Clusters from 2D Observables in MillenniumTNG with Machine Learning

Ana Maria Delgado[1], Michelle Ntampaka[1,2], Sownak Bose[3], Fulvio Ferlito[4], Boryana Hadzhiyska[5,6], Lars Hernquist[7], John Soltis[2], John F. Wu[1,2], Mikaeel Yunus[1], and John ZuHone[7]
[1] Department of Physics & Astronomy, Johns Hopkins University, 3400 N. Charles Street, Baltimore, MD 21218, USA; ana_maria.delgado@cfa.harvard.edu
[2] Space Telescope Science Institute, Baltimore, MD 21218, USA
[3] Institute for Computational Cosmology, Department of Physics, Durham University, South Road, Durham, DH1 3LE, UK
[4] Max-Planck-Institut für Astrophysik, Karl-Schwarzschild-Str. 1, 85748, Garching, Germany
[5] Institute of Astronomy, Madingley Road, Cambridge, CB3 0HA, UK
[6] Kavli Institute for Cosmology Cambridge, Madingley Road, Cambridge, CB3 0HA, UK
[7] Harvard-Smithsonian Center for Astrophysics, 60 Garden Street, Cambridge, MA 02138, USA
*Received 2025 November 25; revised 2026 March 24; accepted 2026 April 4; published 2026 May 4*

## Abstract

Properties of massive galaxy clusters, such as mass abundance and concentration, are sensitive to cosmology, making cluster statistics a powerful tool for cosmological studies. However, favoring a more simplified, spherically symmetric model for galaxy clusters can lead to biases in the estimates of cluster properties. In this work, we present a deep learning approach for estimating the triaxiality and orientations of massive galaxy clusters (those with masses $\gtrsim 10^{14}\,M_\odot h^{-1}$) from 2D observables. We utilize the flagship hydrodynamical volume of the suite of cosmological-hydrodynamical MillenniumTNG (MTNG) simulations as our ground truth. Our model combines the feature extracting power of a convolutional neural network and the message passing power of a graph neural network in a multimodal, fusion network. Our model is able to extract 3D geometry information from 2D idealized cluster multiwavelength images (soft X-ray, medium X-ray, hard X-ray, and tSZ effect) and mathematical graph representations of 2D cluster member observables (line-of-sight radial velocities, 2D projected positions and *V*-band luminosities). Our network improves cluster geometry estimation in MTNG by 30% compared to assuming spherical symmetry. We report an $R^2 = 0.85$ regression score for estimating the major axis length of triaxial clusters and correctly classifying 71% of prolate clusters with elongated orientations along our line of sight.

## 1. Introduction

Galaxy clusters are the largest gravitationally bound objects in the Universe. They are composed of dozens to thousands of galaxies and a hot intracluster medium (ICM) embedded in a dark matter halo. Massive clusters (those with masses $\gtrsim 10^{14}\,M_\odot h^{-1}$) contain most of the dark matter content of the Universe, and their number abundance is sensitive to the underlying cosmological parameters, particularly the amplitude of density fluctuations $\sigma_8$ and the matter density $\Omega_m$. Because of this, they can serve as valuable cosmological probes (for a review, see S. W. Allen et al. 2011; J. L. Tinker et al. 2012).

Cluster formation is highly dependent on the large-scale structure (N. A. Bahcall 1988; M. Einasto et al. 2001; X. Yang et al. 2005; C. Papovich 2008; J. P. Willis et al. 2013). Clusters reside in nodes of the cosmic web and are generally aligned along a main filament that feeds matter into the cluster. They form hierarchically (S. Dodelson & F. Schmidt 2020), growing as they accrete dark matter, gas, galaxies, and other smaller galaxy clusters from their local environment.

Presently, clusters can be identified through observations across the electromagnetic spectrum closely probing the different physical processes acting together during their formation and evolution. The hot, ionized gas of the ICM emits X-rays via bremsstrahlung, collisional excitation, recombination radiation, and two-photon emission processes. Clusters are also observable through inverse-Compton scattering of photons from the cosmic microwave background (CMB) off the electrons in the ICM, an effect known as the thermal Sunyaev-Zeldovich (tSZ) effect (R. A. Sunyaev & Y. B. Zeldovich 1972). The depth of the gravitational potential well of a cluster can be further probed in the optical and infrared via the line-of-sight (LoS) radial velocities of its galaxies, a method that is less affected by the complex baryonic physics influencing the ICM (A. Saro et al. 2013). Weak lensing (mainly in the optical and near-infrared) allows one to detect galaxy clusters without making any assumptions about the baryonic physics, richness (galaxy count), or dynamical state of the cluster, often serving to calibrate cluster mass estimates derived from X-ray or tSZ observations. Although this breadth of methods enables follow-up studies and cross-correlations, all techniques share the common limitation of inferring 3D cluster structure from projected 2D observables.

In cosmological analysis, massive clusters are often treated as spherically symmetric. However, this simplifying assumption ignores the complex morphology and assembly history of clusters. Simulations suggest that the formation history of galaxy clusters substantially influences their morphology (K.-W. Wong & J. E. Taylor 2012; B. Diemer & A. V. Kravtsov 2014; F. Jiang & F. C. van den Bosch 2016; E. T. Lau et al. 2021).

*N*-body simulations show that dark matter halos are predominantly triaxial (J. Dubinski & R. G. Carlberg 1991; Y. P. Jing & Y. Suto 2002; B. Allgood et al. 2006; G. Despali et al. 2014), and massive clusters that are still undergoing collapse are particularly nonspherical as a consequence of nonequilibrium effects. Simulation-based studies have shown that spherical assumptions can introduce a scatter of $\sim$20% in weak lensing mass estimates (M. Meneghetti et al. 2010; M. R. Becker & A. V. Kravtsov 2011; R. Herbonnet et al. 2019). Furthermore, triaxial geometry affects observables, such as X-ray emission symmetry and brightness concentration (E. Rasia et al. 2013), and it is the primary contributor to the characteristic $\sim$35% scatter in dispersion-based mass estimates (A. Saro et al. 2013).

Assumptions of spherical symmetry also introduce systematic biases in cluster mass estimation and cluster selection. Cluster orientation in particular becomes a problem for clusters that are highly prolate along our LoS (V. L. Corless & L. J. King 2007, 2008; F. Feroz & M. P. Hobson 2012) or highly oblate across the plane of the sky (F. Feroz & M. P. Hobson 2012). Although weak lensing studies utilize large samples of clusters to try to reduce orientation bias, observational studies have shown scatter with cluster mass in the selection property of clusters across multiple wavelengths, $\sim$40% for X-ray luminosity, $\sim$20%–45% for optical richness, and $\sim$20% in the SZ ( E. Rozo et al. 2015; A. B. Mantz et al. 2016; R. Murata et al. 2018; A. Farahi et al. 2019), where the scatter in richness and SZ signal are expected to be particularly sensitive to cluster orientation (R. E. Angulo et al. 2012; D. A. Buote & P. J. Humphrey 2012; J. P. Dietrich et al. 2014). Furthermore, a study by M. Sereno et al. (2012) found that assuming a prolate ellipsoidal gas distribution and a highly ellipsoidal halo can resolve disagreements between lensing and X-ray mass estimates that can reach 30%–40% in the inner regions of a cluster. These biases underscore the need for advanced triaxial modeling as a core component of a modern approach to using galaxy clusters as a cosmological probe.

Observational studies that model the triaxial mass distribution of clusters typically adopt a Bayesian framework to marginalize over orientation angles and shape parameters, resulting in constraints such as minor-to-major axis ratios (J. Lee & Y. Suto 2004; M. Limousin et al. 2013; M. Sereno et al. 2017, 2018; I. N. Chiu et al. 2018; K. Umetsu et al. 2018; J. Kim et al. 2024; A. Gavidia et al. 2025). However, the success of these techniques relies on the flexibility of the underlying models to capture the true mass distribution of the cluster.

Machine learning (ML) methods, on the other hand, provide the flexibility needed to capture the nuance of cluster structure. ML has become a widely used and powerful tool for cosmological analysis (S. B. Green et al. 2019; M. Ho et al. 2019, 2021, 2023; Z. Yan et al. 2020; P. Villanueva-Domingo et al. 2021; T. L. Makinen et al. 2022; H. Shao et al. 2023). Its ability to discover patterns in data and capture high-dimensional correlations makes it an important addition for analyzing statistical samples of cosmological probes. Combined with the use of high-resolution hydrodynamical simulations to create heterogeneous statistical datasets analogous to real survey data, we can train an ML pipeline to predict the triaxial shapes directly and refine our theoretical models.

In this work, we present a novel fusion deep learning network, which we henceforth refer to as a hybrid neural network (hNN), for estimating the triaxial geometry of galaxy clusters from heterogeneous, 2D observables obtained from a cosmological hydrodynamical simulation with a dense sampling of massive clusters. For clarity, we refer separately to "triaxiality," which implies the three semiaxes lengths of a cluster, and "orientation," which refers solely to the orientation of the principal axis of the cluster. Our novel hNN leverages the feature extraction capability of convolutional neural networks (CNNs) to probe the ICM via X-ray and tSZ images, combined with the connectivity power of graph neural networks (GNNs) to probe the gravitational potential well via LoS galaxy radial velocities in clusters, to directly estimate a triaxial geometry. Our work underscores the necessity of moving beyond spherical approximations in the era of precision cosmology, where systematic uncertainties increasingly dominate over statistical errors in cosmological studies utilizing galaxy clusters as probes. The organization of this paper is as follows: In Section 2 we describe the simulation used in this study, summarize our cluster selection, define our cluster geometry variables, describe our idealized 2D multiwavelength images, and outline our hNN pipeline. In Section 3 we report the results of our various ML tasks, and in Section 4 we provide a summary and discussion of our work.

## 2. Methods

### 2.1. The MillenniumTNG Simulation

In order to build a cluster catalog large enough to train a machine learning algorithm, with resolution capable of forming realistic galaxies and modeling the gas physics of the ICM, we employ the highest-resolution full-physics simulation from the MillenniumTNG project, henceforth MTNG. The MTNG volume has a resolution sufficient to describe the physical processes of galaxies, providing us with the ability to make multiband mock observables that track the internal dynamics of galaxy clusters. The galaxy clusters in MTNG agree well with those found in Planck and the Sloan Digital Sky Survey (SDSS)-8 RedMaPPer richness catalog in observational space (R. Pakmor et al. 2023). Moreover, the MTNG volume is large enough to provide an adequate sample of massive clusters with which to perform our study.

MTNG is a periodic box with comoving volume $(L_{\rm box} = 500\,h^{-1}{\rm Mpc} \approx 740\,{\rm Mpc})^3$ using $4320^3$ dark matter particles with mass $1.7 \times 10^8\,M_\odot$, and $4320^3$ gas cells, each with an initial mass of $3.1 \times 10^7\,M_\odot$. The physics model of MTNG is based on the IllustrisTNG (F. Marinacci et al. 2018; J. P. Naiman et al. 2018; D. Nelson et al. 2018; A. Pillepich et al. 2018a; V. Springel et al. 2018) galaxy formation physics model (R. Weinberger et al. 2017; A. Pillepich et al. 2018b; D. Nelson et al. 2019a, 2019b; A. Pillepich et al. 2019), which has been shown to produce realistic galaxy populations on cosmological scales. All cosmological parameters are given by Planck Collaboration et al. (2016a) as in IllustrisTNG: $\Omega_0 = 0.3089$ (total matter density), $\Omega_b = 0.0489$ (baryon density), $\Omega_\Lambda = 0.6911$ (dark-energy density), $H_0 = 67.74\,{\rm km\,s^{-1}\,Mpc^{-1}}$ (Hubble parameter), and $\sigma_8 = 0.8159$ (linear rms density fluctuations in a sphere of radius $8\,h^{-1}$ Mpc at $z = 0.0$). Initial conditions were generated at $z = 63$ with second-order Lagrangian perturbation theory using GADGET-4 (V. Springel et al. 2021). All full-physics simulations in the suite were run

**Table 1**
Brief Description of Cluster Geometry Variables for Reference

| Parameter | Equation | Description |
|---|---|---|
| $\lvert e_{1_z} \rvert$ | ⋯ | Describes principal axis orientation. This is the absolute value of the $z$-component of the principal eigenvector. |
| $\epsilon_+$ | (3) | 2D projected alignment along $x$-axis of FoV |
| $\epsilon_\times$ | (4) | 2D projected alignment along $y$-axis of FoV |
| $\epsilon_{\rm FoV}$ | (6) | 2D projected ellipticity, where $\epsilon_{\rm FoV} = 0$ is a circle, while $\epsilon_{\rm FoV} = 1$ is a line |
| $\Delta_\epsilon$ | (7) | Describes the difference between the 3D shape (minor to major axis ratio) and the observed 2D ellipticity, where $\Delta_\epsilon = -1$ is flat and elongated while $\Delta_\epsilon = 1$ is a perfect sphere. |

**Note.** Each variable is described in detail in Section 2.2.1.

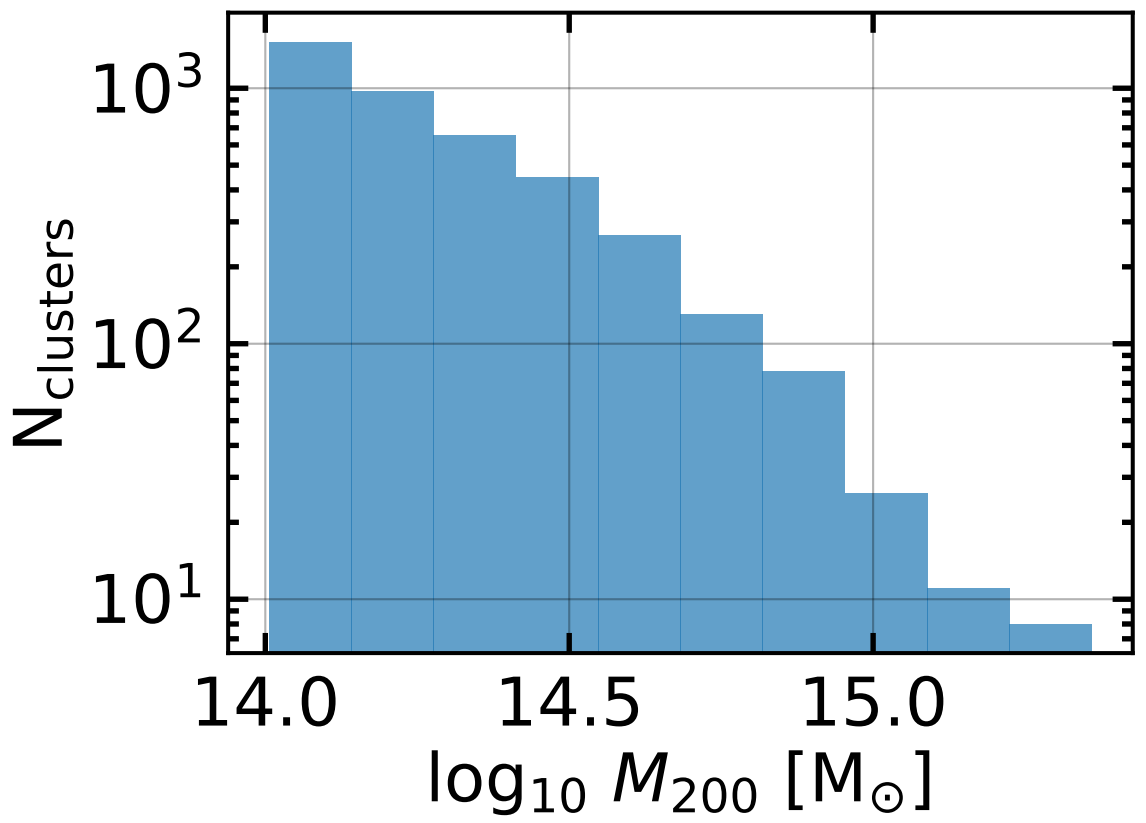

**Figure 1.** Histogram of MTNG cluster mass. We define massive clusters as those having of $M_{200} > 10^{14} M_\odot$, yielding 4117 massive clusters in MTNG. Most cluster masses range from $\log_{10}(14.0 - 14.5) M_\odot$. For machine learning applications, we must ensure that low-frequency clusters are properly distributed across train/validation/test splits.

using the AREPO code (V. Springel 2010; R. Weinberger et al. 2020). We note that to reduce the memory requirements of the MTNG simulation, minor modifications were made to the physics model as compared to IllustrisTNG; namely, magnetic fields and following individual metal species needed to be disabled in MillenniumTNG. These changes, particularly removing the magnetic fields, have a slight but measurable impact on galaxy properties (R. Pakmor et al. 2017).

### 2.2. Cluster Selection

We select massive clusters from the MTNG group catalog with total mass $M_{200} > 10^{14} M_\odot$ with respect to (w.r.t.) the critical density of the Universe ($M_{200c}$), yielding a sample of 4117 clusters at $z = 0.0$. The histogram of cluster masses in Figure 1 reveals that our mass distribution is quite imbalanced. In fact, only 34 clusters in our catalog have masses greater than $\log_{10} M_{200} = 15 M_\odot$, with the most massive cluster having a mass of $\log_{10} M_{200} = 15.35 M_\odot$. For machine learning applications, it is important to ensure that low-frequency clusters are properly distributed across train/validation/test splits such that each data split contains at least two clusters with mass greater than $\log_{10} M_{200} = 15 M_\odot$.

#### 2.2.1. Shape and Orientation

Although mass serves as the primary descriptor in most cluster studies, our work aims to establish cluster ellipticity as a powerful descriptor that captures important information that is often a source of systematic error in observational mass estimates. The cluster geometry variables described in this section are summarized Table 1. Our methods for measuring cluster ellipticity and orientation are as follows.

We determine the triaxial semimajor axis lengths of a cluster from the positions of its subhalos. The second-moment tensor can be defined as

$$q_{ij} = \frac{1}{W} \sum_n (\boldsymbol{x}_n - \bar{\boldsymbol{x}})_i (\boldsymbol{x}_n - \bar{\boldsymbol{x}})_j w_n, \quad (1)$$

where $W$ is the total weighted value for the distribution, $\boldsymbol{x}_n$ is the position vector of the subhalo, $n$, and $w_n$ is the weighted value for the subhalo. For this study, we adopt uniform subhalo weighting, for each subhalo $w_n = 1$. We obtain the semiaxis lengths of the cluster from the square root of the ordered eigenvalues of the second-moment tensor, which we will refer to as $\lambda_a$, $\lambda_b$, and $\lambda_c$, respectively (where $\lambda_a > \lambda_b > \lambda_c$). We compared the three axis ratios derived from uniform weighted subhalos to those from using the total mass weighted matter particles in each cluster and found consistent measurements, with the mean difference $\overline{\Delta} \approx 0.05 \pm 0.04$ While many physically motivated weighting schemes exist (e.g., luminosity weighting, stellar mass weighting, gas mass weighting), we expect some dependence on this choice since tracers that are more spatially biased to the underlying distribution of matter can yield different axis ratios. For example, luminosity weighted subhalo positions yield $\overline{\Delta} \approx 0.14 \pm 0.10$ compared to the total mass weighted particle positions, statistically significantly larger than our chosen scheme.

We obtain the orientation of the three axes from the corresponding sorted eigenvectors of the second-moment tensor as $e_1$, $e_2$, and $e_3$. For our study, we are most interested in the orientation of a cluster along our chosen LoS: the Cartesian $z$-axis of the simulation box. Thus, the value of the $z$-component of the principal eigenvector, $e_1$, is most informative for our purposes. Additionally, we only consider the absolute alignment of the cluster (the positive/negative direction is irrelevant) and henceforth refer to this parameter as $\lvert e_{1_z} \rvert$.

From the 3D moments of the position distribution, one can further determine the projected moments, for example, along the Cartesian $z$-axis of the simulation box, which we adopt as our default projection direction. We define the projected, complex ellipticity, $\epsilon$, as follows:

$$\epsilon = \epsilon_+ + \mathrm{i}\epsilon_\times = \frac{q_{xx} - q_{yy}}{q_{xx} + q_{yy}} + \mathrm{i}\frac{2q_{xy}}{q_{xx} + q_{yy}} \quad (2)$$

where $\epsilon_+$ refers to the elongation of the cluster along the $x$–$y$ plane, while $\epsilon_\times$ measures the elongation at a 45° angle from the axes. $\epsilon$ transforms under rotation as a traceless, symmetric $2 \times 2$ tensor. For an angle $\theta$, measuring the angle between the principal axis and the $x$-axis of our field of view (FoV), epsilon changes as $\epsilon' = \epsilon \exp(2i\theta)$, giving

$$\epsilon_+ = \epsilon \cos(2\theta) \quad (3)$$

$$\epsilon_\times = \epsilon \sin(2\theta). \quad (4)$$

These variables provide orientation information, such that $\epsilon_+ > 0$ corresponds to a cluster whose projected ellipticity is aligned along the $x$-axis of the FoV, $\epsilon_+ < 0$ corresponds to alignment along the $y$-axis, $\epsilon_\times > 0$ corresponds to diagonal alignment along $x = y$, and $\epsilon_\times < 0$ corresponds to alignment along $x = -y$. These values can be estimated observationally using the light-weighted calculation of the second-moment tensor. In this study, we use the estimation of $\epsilon_+$ and $\epsilon_x$ as a "sanity check" and to assess the quality of the hNN predictions against quantities that can be measured directly in observations.

We can similarly utilize only the 2D positions of the dark matter particles for $\boldsymbol{x}_n$ in Equation (1) in order to obtain axis lengths for a projected ellipse across our FoV, $\lambda_{a_{2D}}$ and $\lambda_{b_{2D}}$. With both 2D and 3D semiaxes lengths, we define a cluster shape and orientation parameter in the following manner.

For 3D intrinsic shape information, we employ the 3D axis ratios of the minor to major semiaxes, $\frac{\lambda_c}{\lambda_a}$. This is useful because if the ratio is equal to 1, the cluster is a sphere, since in that case $\lambda_c = \lambda_a$, and thus, it is logical to conclude that $\lambda_c = \lambda_a = \lambda_b$, as the intermediate semiaxis must fall between the minor and major semiaxes. If, however, $\frac{\lambda_c}{\lambda_a} \to 0$, the cluster is either a highly prolate ellipsoid (a "cigar" shape) if $\lambda_b \approx \lambda_c$ or a highly oblate ellipsoid (a "pancake" shape) if $\lambda_b \approx \lambda_a$.

Additionally, for 2D FoV ellipticity, we define:

$$q = \frac{\lambda_{b_{2D}}}{\lambda_{a_{2D}}} \quad (5)$$

$$\epsilon_{\rm fov} = \frac{1 - q^2}{1 + q^2}, \quad (6)$$

where $q$ is the 2D axis ratio, and $\epsilon_{\rm fov}$ describes the 2D ellipticity of the cluster across the FoV, and has the value $\epsilon_{\rm fov} = 1$ for a line, and of $\epsilon_{\rm fov} = 0$ for a circle. We note that our definitions map to conventions commonly used in observational triaxial cluster analyses. Specifically, our 2D axis ratio, $q$, corresponds to the observed projected axis ratio, $q_p$, and our ellipticity across the FoV, $\epsilon_{FoV}$, is analogous to the linear form of $\epsilon = 1 - q_p$ used in observational studies (e.g., M. Sereno et al. 2012; J. Kim et al. 2024), where likewise $\epsilon = 1$ is a line, and $\epsilon = 0$ is a circle. The main difference is that simulation-based analyses measure these values directly from the second-moment tensors while observational studies fit projected surface brightness isophotes. Figure 2 illustrates $\epsilon_{\rm FoV}$ measurements for randomly selected clusters from our MTNG cluster catalog. The distribution of dark matter is shown in blue, gas in magenta, and subhalos weighted by $v$-band luminosity are shown in orange. We can see that gas and subhalos closely trace the dark matter distribution in the clusters. The 2D ellipticities of clusters, $\epsilon_{\rm fov}$, are drawn as ellipses. The blue ellipse corresponds to $\epsilon_{\rm FoV}$ calculated with equal weighting of Equation (1) with dark matter particles; similarly, the magenta ellipse is that for the gas particles, the orange ellipse is done weighted by the luminosity of the subhalos in the $v$ band, and the green ellipse is obtained by using all of the subhalos in the cluster and giving each an equal weight of 1. For the purpose of this study, we utilize $\epsilon_{FoV}$ as obtained by the equal weighted subhalo positions (the green ellipse).

Inferring 3D cluster morphology from 2D projections is inherently degenerate. Therefore, in order to identify and isolate morphologically ambiguous systems from projection effects, we classify cluster morphology using the following ad hoc scheme:

$$\Delta_\epsilon = \frac{\lambda_c}{\lambda_a} - \epsilon_{\rm fov}., \quad (7)$$

where $\frac{\lambda_c}{\lambda_a}$ is the intrinsic minor to major axis ratio, and $\epsilon_{FoV}$ is the projected ellipticity across our FoV. $\Delta_\epsilon \approx 1$ indicates a true sphere (round both intrinsically and in projection), while $\Delta_\epsilon \approx -1$ indicates a highly elongated system that also appears as a flattened ellipsoid in projection. The intermediate range, $\Delta_\epsilon \approx 0$, represents a morphologically degenerate region where many different configurations (varying triaxiality, axis ratios, and orientation) can produce similar $\Delta_\epsilon$ values, making reliable shape inference challenging. Figure 3 illustrates these different morphological regions.

To isolate the ambiguous morphological population, we empirically define boundaries at $\Delta_\epsilon \pm 0.2$ based on visual inspection of clusters and their distributions of projected ellipticity. Clusters with $\Delta_\epsilon > 0.2$ are classified as "spheroids," while those with $\Delta_\epsilon < 0.2$ are "ellipsoids." These boundaries provide clean morphological separation, leaving the degenerate region, $0.2 \leqslant \Delta_\epsilon \leqslant 0.2$, with a population of clusters whose morphology cannot be reliably determined from projection alone. Within this degenerate region lies a unique subset of particular interest to this study: prolate ellipsoids oriented along our LoS that appear deceptively circular in projection despite significant intrinsic elongation. We therefore isolate these rare objects further by requiring $|e_{1_z}| \geqslant 0.75$ and refer to them as "false spheres."

Figure 4 illustrates the nature of the morphological degeneracy, and highlights the utility of the $\Delta_\epsilon$ boundaries. Spheroids (orange) concentrate at low $\epsilon_{\rm FoV}$ (appearing round), while ellipsoids (green) concentrate at high $\epsilon_{FoV}$ (appearing elongated). Critically, the intermediate $\epsilon_{\rm FoV}$ range is dominated by the degenerate region of $\Delta_\epsilon$ (black hatching) where intrinsic 3D morphology cannot be reliably determined from projected 2D appearance alone. This region contains false spheres (red) and genuinely ambiguous systems (blue).

The histogram in Figure 5 illustrates the distribution of $|e_{1_z}|$ values for samples of different cluster morphologies. We highlight that false spheres (shown in red) are rare and concentrate at high $|e_{1_z}| \geqslant 0.75$ range, as expected for LoS systems. We remind the reader that the difference between ellipsoids and false spheres, as we have defined them, is that ellipsoids appear elliptical across our FoV, while the false spheres appear circular. The ambiguous population (systems in the degenerate region that are not strongly aligned along our LoS, shown in blue) includes both mildly ellipsoidal and spheroidal morphologies that cannot be confidently classified and spreads across the remaining orientation distributions. A complete population of all ellipsoids (identified ellipsoids, shown in green, plus false spheres plus ellipsoidal members of ambiguous) and of all spheroids (identified spheroids, shown in orange, plus spheroidal members of ambiguous) would each exhibit nearly isotropic orientation distributions. The apparent concentration of identified spheroids at high $|e_{1_z}|$ is an artifact of measuring discrete particle positions with finite moment tensors, and we emphasize that eigenvector orientations are physically meaningless in spherical systems. Indeed, the high abundance of spheroids and their full range of $|e_{1_z}|$ create noise that make the task of identifying both triaxiality and orientation challenging. We note that the full sample of clusters (gray) exhibits an approximately uniform distribution consistent with isotropic orientations.

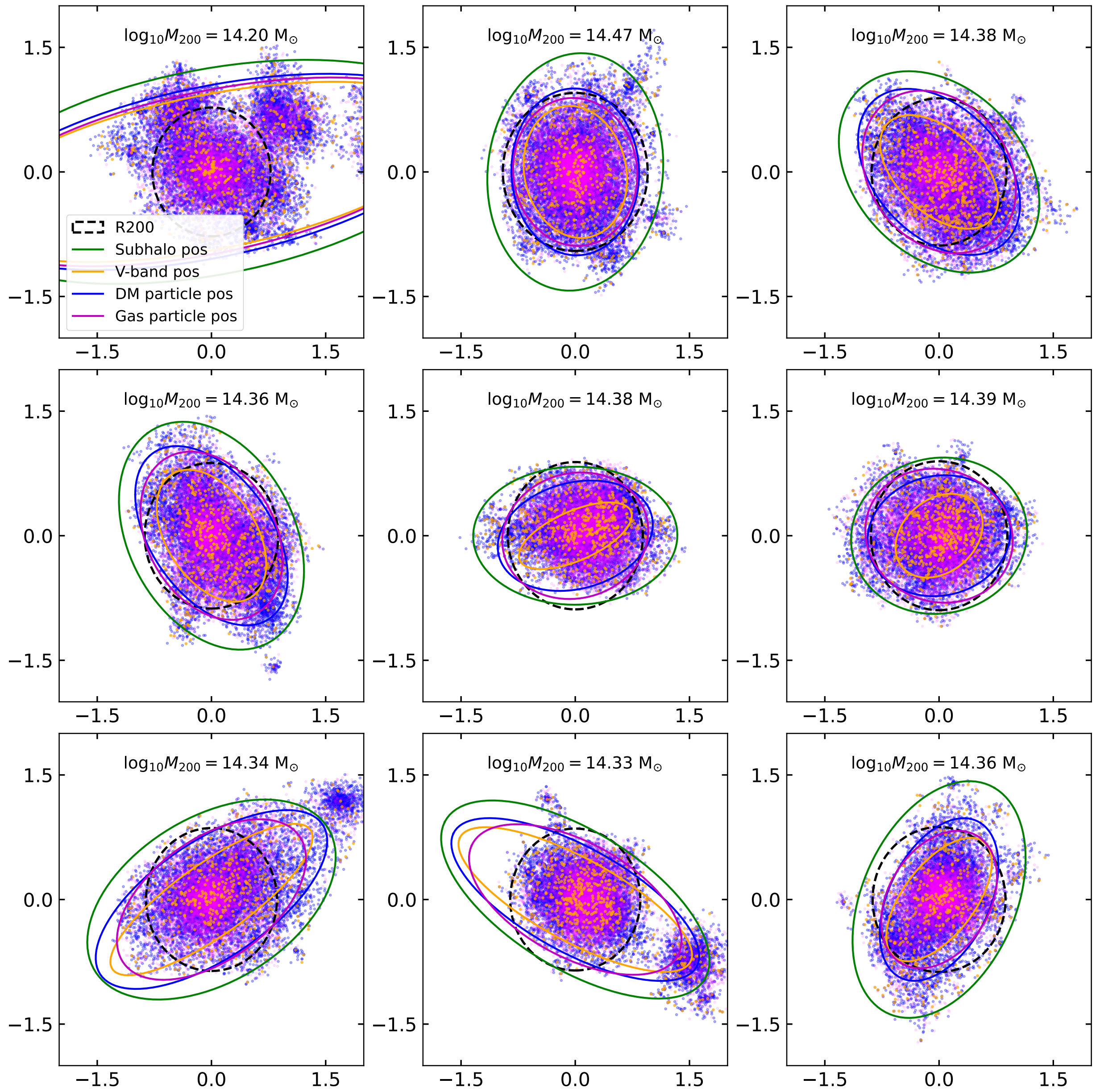

**Figure 2.** Visualization of massive MTNG clusters. All *x*- and *y*-axes are in units of $h^{-1}$Mpc. In each panel, dark matter particles (sampled at 1-in-100) are shown in blue, gas particles (sampled at 1-in-100) are shown in magenta, and subhalos are shown in orange. Corresponding ellipses are drawn for the $\epsilon_{\rm FoV}$ of the positions. Blue ellipses correspond to $\epsilon_{\rm FoV}$ measured from the dark matter positions weighted by particle mass, magenta ellipses correspond to those of gas particle positions weighted by particle mass, and the orange ellipses correspond to those of *V*-band luminosity weighted subhalo positions. We additionally show green ellipses that correspond to the $\epsilon_{\rm FoV}$ measured using all subhalos given equal weighting. The black dashed circle shows where $R_{200}$ (radius within which the mean density of the host halo equals 200 times the critical density of the Universe) lies from the center of the cluster. For this work, we utilize $\epsilon_{\rm FoV}$ of all subhalo positions, as shown in the green ellipses.

Lastly, we classify cluster orientation into three categories: alignment across our FoV with $|e_{1_z}| \in [0.0, 0.3)$ is classified as "FoV," systems with $|e_{1_z}| \in [0.3, 0.7)$ are classified as "partial" as they are partially aligned along our LoS, and csystems aligned along our LoS with $|e_{1_z}| \in [0.7, 1.0]$ are classified as "LoS." These boundaries approximate equal bins (which would be *t* 0.33 and 0.677), rounded for cleaner thresholds.

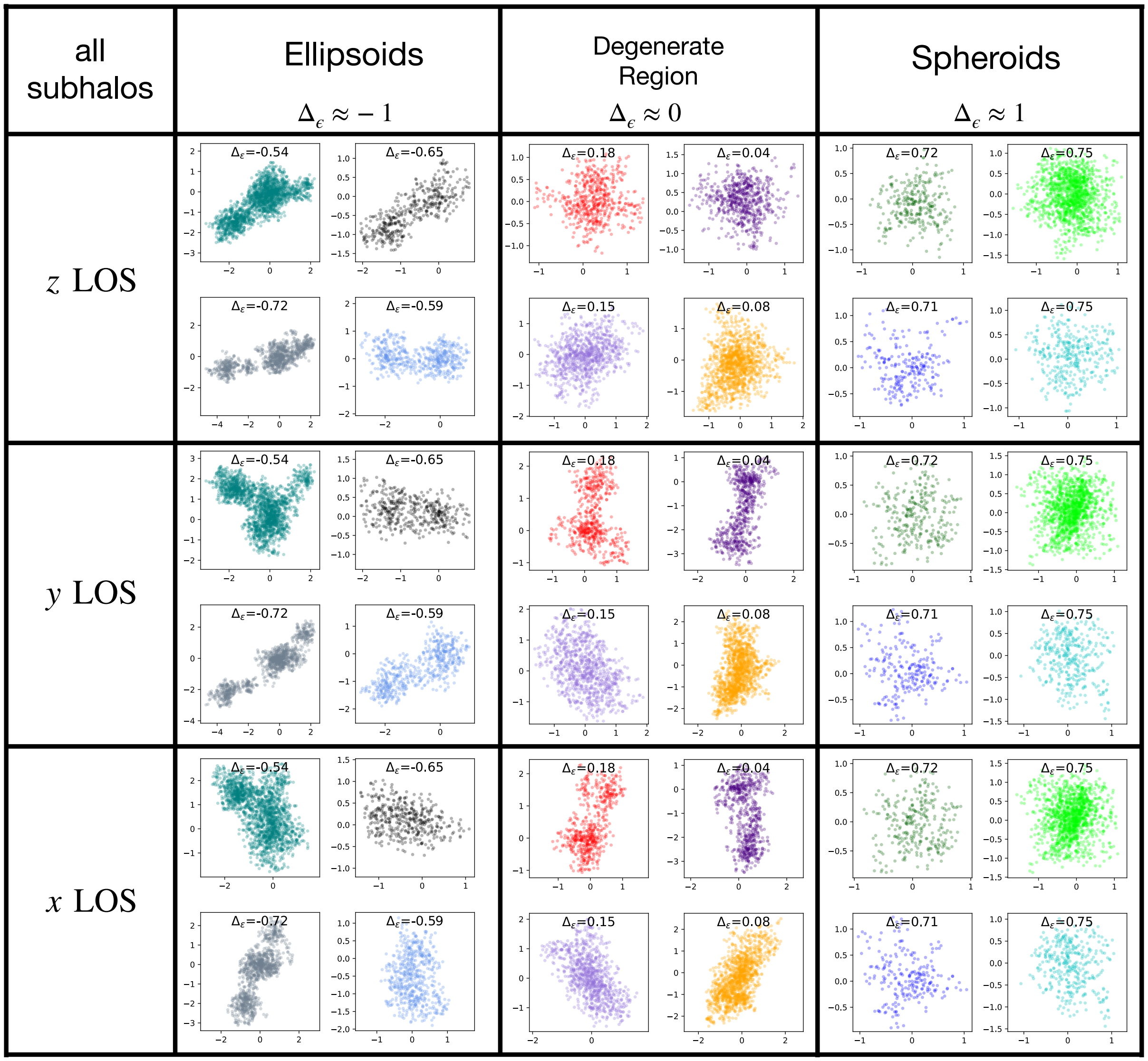

**Figure 3.** Cluster subhalo positions as they would appear projected onto the sky from various viewing angles ($z$-LoS, $y$-Los, $x$-LoS), illustrating the difference between a cluster's projected shape and intrinsic shape. Each column contains four different clusters grouped in the same morphological region as determined by $\Delta_\epsilon$. We draw the reader's attention to the middle column containing clusters in the degenerate region where one can't reliably infer the true 3D shape from the 2D projection. This region contains clusters that have a significantly elongated projected morphology across either or both of the $y$-LoS and $x$-LoS views compared to our chosen $z$-LoS view. A special subgroup in this degenerate region is a population we refer to as "false spheres," which have a circular projection perpendicular to our LoS, but are significantly elongated along our LoS. Ellipsoids in the left column have elongated projected morphologies in all views, and true spheroids in the right column have circular projected morphologies in all views. The axes in each subpanel have units of $h^{-1}$ Mpc.

The boundaries for LoS-aligned systems encompass our false sphere criterion ($|e_{1_z}| \geqslant 0.75$), making false spheres the most extreme subset of LoS-aligned clusters.

### 2.2.2. Idealized X-ray and tSZ Observations

Our idealized X-ray and tSZ observations of the MTNG clusters are similar to those in M. Ntampaka et al. (2019) and J. Soltis et al. (2025), respectively, which we will briefly describe here. The idealized X-ray observations are produced using the `pyXSIM`[8] package (J. A. ZuHone et al. 2014). The X-ray photon intensity (in counts $s^{-1}$ $cm^{-2}$ $arcsec^{-2}$) for each gas cell in the simulation is derived using an APEC emission model (A. R. Foster et al. 2012), taking the density, temperature, and metallicity data of the cells in each cluster as input, and assuming a redshift of $z = 0.05$. Only gas cells with a temperature $T > 3 \times 10^5$ K and a gas density $\rho_g < 5 \times 10^{-25}$ g cm$^{-3}$ that are not forming stars are used in the construction of the intensity maps, which are projected along our chosen LoS ($z$-axis). Each X-ray observation is separated into three bands: soft (0.5–1.2 keV), medium (1.2–2.0 keV), and hard (2.0–7.0 keV).

For the tSZ observations of the MTNG clusters, we produced projected maps for the Compton-y parameter, $y_{\rm tSZ}$:

$$y_{\rm tSZ} = \int \frac{k_{\rm B} T}{m_e c^2} \sigma_T n_e d\ell, \quad (8)$$

where $n_e$ is the electron number density, and $\ell$ is the path-length along the LoS.

The FoV corresponds to ~8 Mpc across at $z = 0.05$. This is large enough to include the full extent of the tSZ information

[8] http://hea-www.cfa.harvard.edu/~jzuhone/pyxsim/

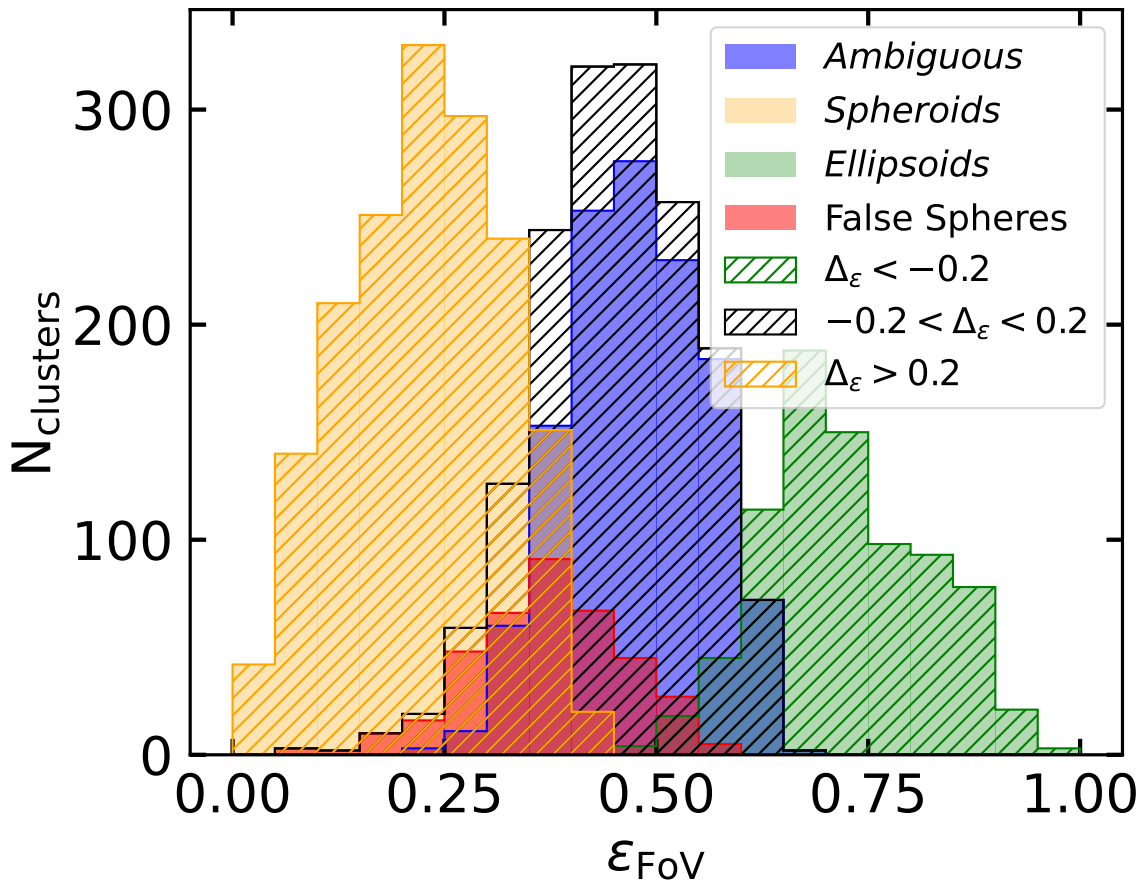

**Figure 4.** A histogram of the projected ellipticity across the plane in the sky, $\epsilon_{\mathrm{FoV}}$, illustrating the nature of morphological degeneracy. Confidently identified ellipsoids (green) and spheroids (orange) have clear boundaries in both $\epsilon_{\mathrm{FoV}}$ and $\Delta_\epsilon$. However, the degenerate region defined by intermediate $\Delta_\epsilon$ values (black hatching), where intrinsic 3D morphology cannot be inferred from projected appearance alone, dominates the intermediate $\epsilon_{\mathrm{FoV}}$ region overlapping in projection both spheroidal and ellipsoidal populations.

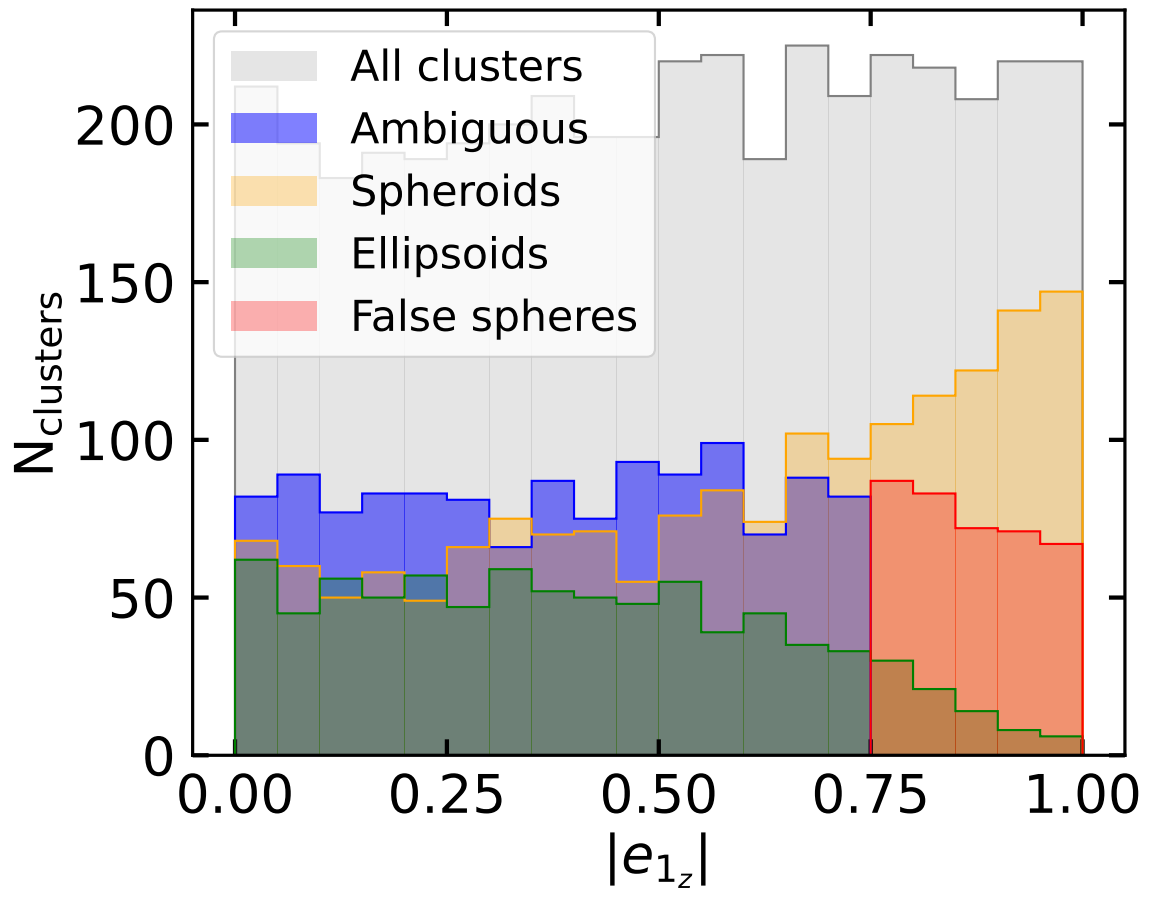

**Figure 5.** A histogram of the distribution of the principle axis orientations, $|e_{1_z}|$, from different morphologies of massive MTNG clusters as defined in Section 2.2.1. While spheroids (orange) and ellipsoids (green) are abundant, false spheres (red) are much rarer and have a limited $|e_{1_z}|$ range. Correctly identifying these false spheres in observations could reduce systematic errors in weak lensing measurements. Ambiguous systems (blue) are not strongly aligned along our LoS, and may include both spheroidal and ellipsoidal clusters. The distribution of all clusters (gray) shows that there is no preferred orientation in the data as a whole.

for most clusters without having an excessive amount of empty space in the X-ray.

Each image has dimensions 128 × 128. They are normalized in the following manner:

$$X' = \tanh\left(\log_{10}(X/\overline{X} + 1)\right), \tag{9}$$

where $X$ is the original 128x128 pixel image, $\overline{X}$ is the mean pixel value for that observation mode across the dataset, and $X'$ is the normalized image. Figure 6 depicts normalized images in the four bands for three clusters: the most massive cluster in our catalog (top row), the median mass cluster (middle row), and the least massive cluster in the catalog (bottom row).

### 2.3. Machine Learning for Cluster Triaxiality and Orientation

As described in Section 1, massive galaxy clusters are powerful cosmological probes, and while it is simpler to summarize these objects by their mass and characterize them as spherically symmetric, accurately modeling the subtleties of their triaxial shape is an essential systematic to address if we are to achieve precision cosmology.

A multiwavelength approach to studying massive clusters has proven beneficial for revealing important substructure and providing insight into the physical processes governing cluster formation. Additionally, machine learning is an obvious tool for extracting high-dimensional correlations and subtle signals encoded in images and complex spatial patterns, making it well suited for extracting triaxiality information from 2D cluster data. Thus, we build on this principle and construct a multimodal neural network architecture, which takes heterogeneous data as inputs and leverages the complementary strengths of different deep learning approaches, each optimized for extracting features from distinct types of 2D cluster observables.

A supervised machine learning algorithm trains a model by providing a subset of data, referred to as the training set, including input variables (often referred to as "features") and output variables (often referred to as "targets"). The goal is to use the training set to learn the connection between the input features and the output variables. During training, the model is optimized by evaluating its predictions on a validation set, typically by way of minimizing a loss function. Finally, the trained model is then used to predict the output variables for a different subset of input features referred to as the test set. We split our cluster catalog (of 4117 total clusters) into an 80/10/10 random split: 80% of the clusters are designated as the training set, 10% as the validation set, and the remaining 10% for the test set.

We use the publicly available `Python` package `Pytorch`[9] to construct a multimodal fusion network, an hNN, by combining a CNN and a GNN along with task-specific multilayer perceptrons (MLPs), described in more detail below.

#### 2.3.1. hNN Architecture Overview

Massive clusters glow brightly in the X-ray and in the tSZ (as described in Section 1 and as visualized in Figure 6) making multiwavelength images of massive clusters a useful probe of cluster morphology w.r.t. the ICM. Additionally, the very nature of a galaxy cluster as a gravitationally bound object with many galaxy members, often quantified through richness studies or through the velocity dispersions of its galaxy members, lends itself to a graph model that can further probe the morphology of a cluster w.r.t. to the projected positions and LoS radial velocities of its members. We therefore develop an hNN architecture that is specifically designed for such inputs, composed of two input branches: a CNN branch ideal for processing four-channel 2D images (soft X-ray, medium X-ray, hard X-ray, and tSZ effect), and a GNN branch ideal for processing 2D cluster observables that can be obtained from optical or near-IR (NIR) surveys in a graph-like structure. Networks combining outputs from a CCN and GNN have been used in a recent study on the galaxy-halo connection (A. J. Larson et al. 2024), though with a network architecture different from the one presented here. This highlights the flexibility with which hNNs can be constructed and for a broad range of applications.

[9] https://pytorch.org/

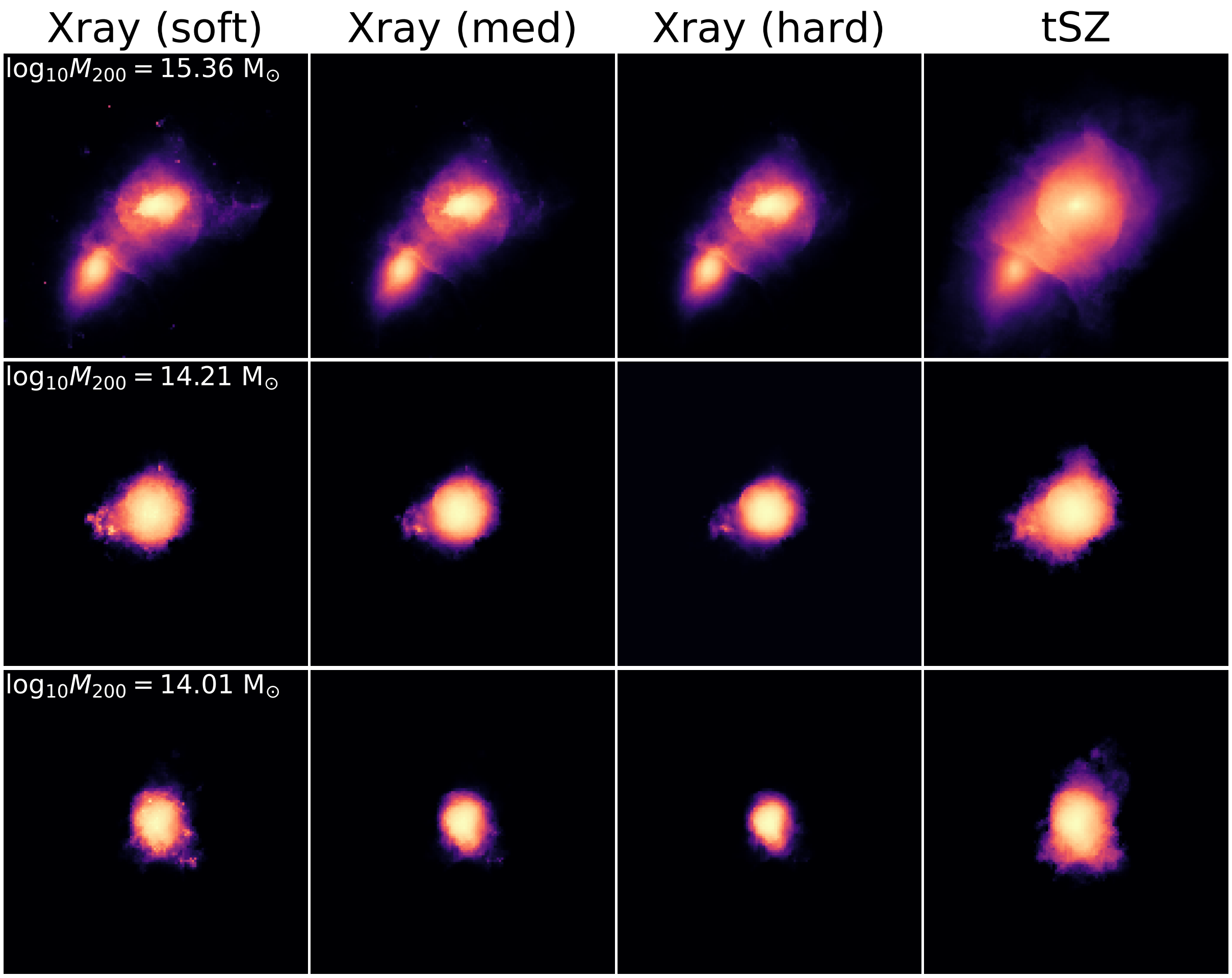

**Figure 6.** Visualization of our four-channel cluster images. Each image is normalized as per Equation (9) with dimensions 128 × 128 pixels. We show images for the most massive cluster in our catalog (top), the median mass cluster (middle), and the least massive cluster (bottom). Differences across the four wavelengths are noticeable by eye.

The CNN is well suited for extracting spatial features from multiwavelength imaging data, while the GNN is optimal for modeling the inherent graph structure of galaxy clusters, where individual galaxies serve as nodes connected by their spatial relationships. The GNN branch performs graph-level predictions by aggregating information across all nodes. Following feature extraction, the latent representations from both branches are combined in a fusion layer where they are first passed through an initial set of fully connected dense neurons and subsequently passed to task-specific MLPs designed to estimate cluster geometry either through regression or classification tasks. Figure 7 provides a visual representation of our pipeline.

### 2.3.2. CNN Branch

For each cluster in our catalog, we concatenate the soft X-ray, medium X-ray, hard X-ray, and tSZ images creating a four-channel input layer where each channel is size 128 × 128 pixels (as shown in Figure 6). We pass this input layer through a hierarchical feature extraction pipeline. The CNN branch comprises three convolutional blocks, each consisting of a 2D convolution. This is followed by batch normalization for training stability, GELU activation to provide smooth, differentiable activation that enhances gradient flow in both the convolutional and GNN components of our architecture, and max pooling for translation invariance and computational efficiency (for a review of CNNs and deep learning, see Y. LeCun et al. 2015).

The first convolutional layer transforms the four-channel input to 16-channel feature maps, while subsequent layers double the channel depth, giving a representation that captures features at multiple scales. The progressive max pooling operations reduce the spatial resolution by a factor of 2 at each stage, resulting in a final feature map of dimension 4 × 16 channel × image size of 8 × 8. These spatially reduced feature maps are then flattened into a 1D vector, creating a compact encoding of the multiwavelength cluster observations that is sent to the feature fusion layer.

### 2.3.3. GNN Branch

We construct a mathematical graph for each cluster in our catalog. Each node in the graph represents a galaxy member of the cluster, which we define as subhalos within the group resolved with a minimum of 50 stellar particles, corresponding to galaxies with $M_\star \gtrsim 1.5 \times 10^9 M_\odot$, comparable to the effective mass range of galaxies in surveys such as SDSS and DESI. The number of resulting galaxy members in our clusters ranges from ∼10–900 members, with most clusters containing approximately 50 members, as shown in Figure 8. An additional benefit of the GNN is that it is inherently capable of accepting input graphs with different numbers of nodes. Unlike, for example, the CNN, where each cluster image must have the same dimensions (128 × 128 pixels), the GNN can accept cluster graphs with different numbers of galaxy members.

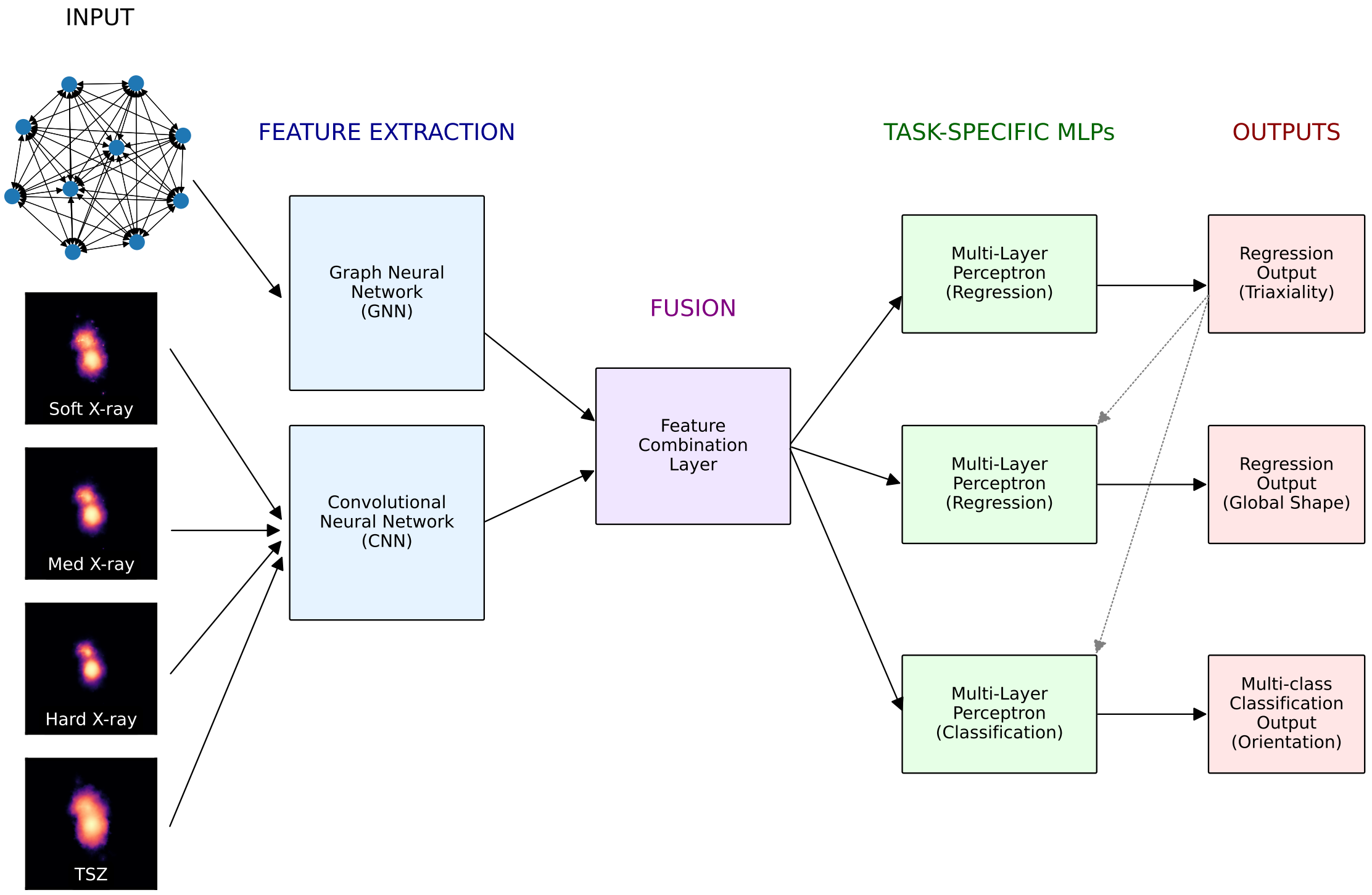

**Figure 7.** Visualization of the hNN pipeline. Our hNN consists of two main branches for feature extraction: a GNN branch designed for graph-level predictions, which takes a graphic representation for each cluster as input, and a CNN branch that takes four-channel 128 × 128 multiwavelength images of each cluster as inputs. The latent representations from each branch are then combined in a fusion layer where further feature extraction occurs. The combined data is then sent to three parallel, task-specific MLPs: one designed for multiparameter regression of the three triaxial semiaxis lengths, another for global cluster shape, and a third MLP designed for multiclass classification, which predicts cluster orientation along our LoS. Skip connections, shown as the dotted gray lines, provide information from the triaxial predictions to the other two tasks.

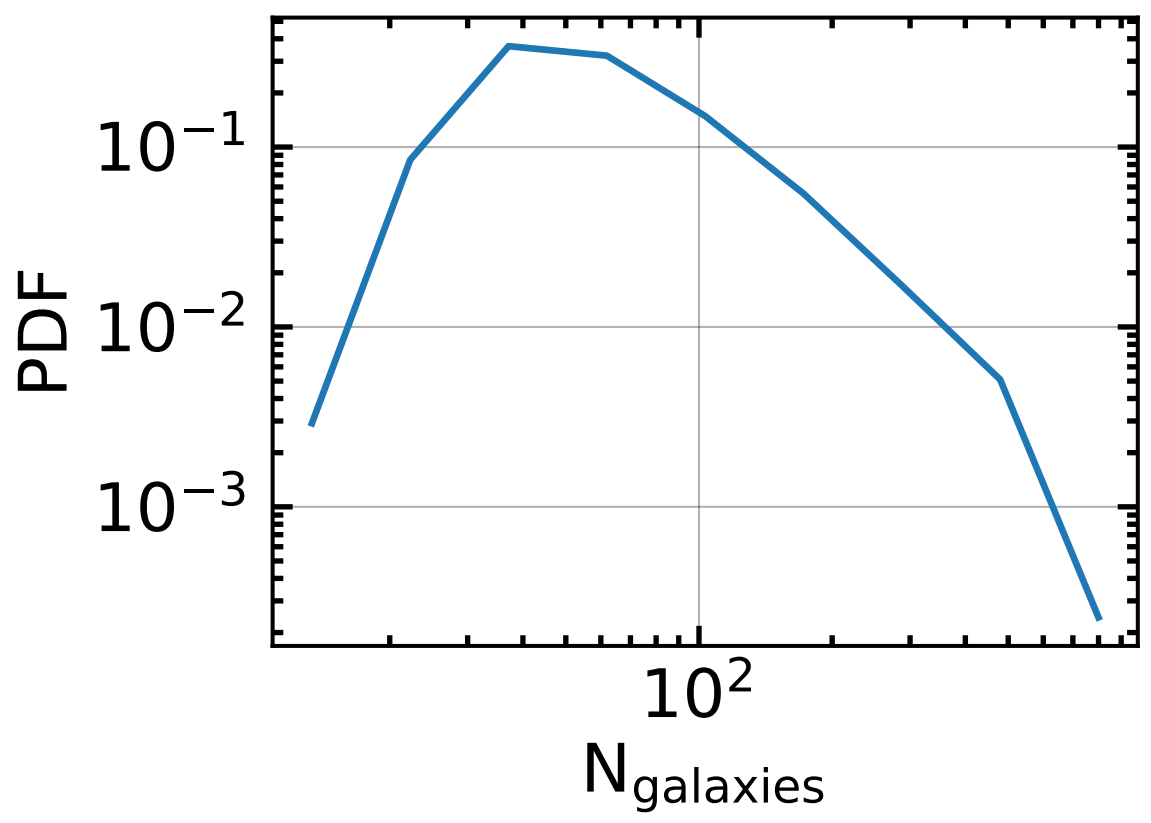

**Figure 8.** Distribution of galaxy members within MTNG clusters. We define galaxies as subhalos resolved with at least 50 stellar particles corresponding to $M_* \gtrsim 1.5 \times 10^9 M_\odot$. Most clusters in our catalog have ~50 galaxies.

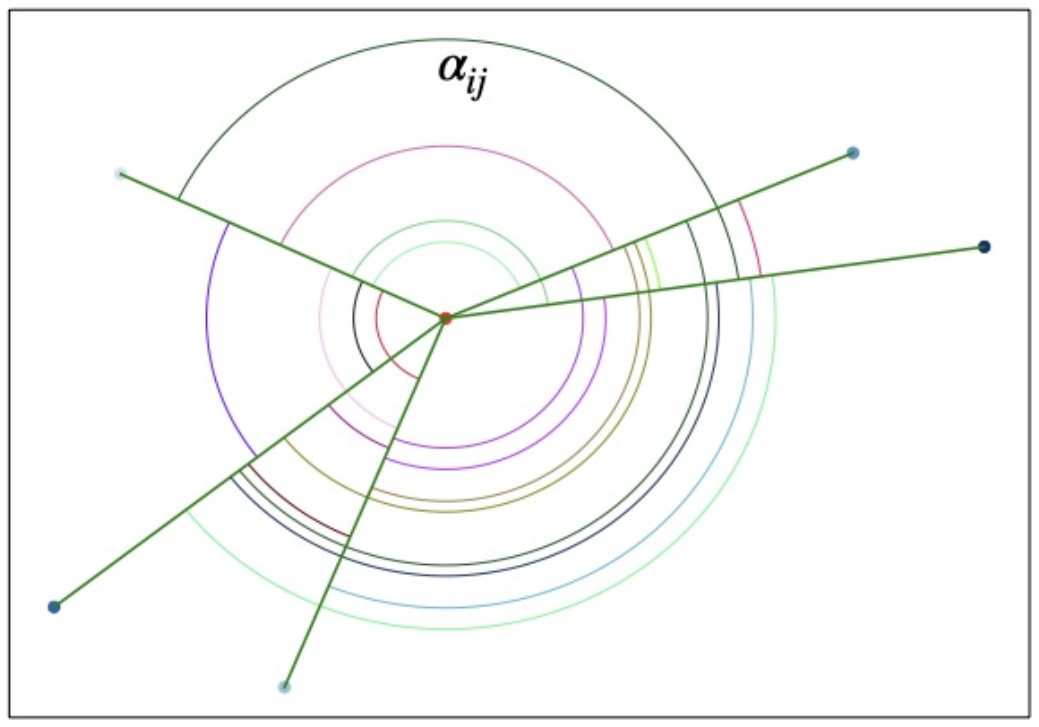

**Figure 9.** A toy illustration of a graph with edge features angle $\alpha$. Each node represents a galaxy, and angle $\alpha$ is seen as the angle that subtends the central galaxy by a pair of two other galaxies.

Each node is represented by $V_i = (R_{xyi}, v_{radi}, L_{vi})$, which are the relative 2D Euclidean positions across our FoV normalized w.r.t. the central node, the LoS radial velocity, and $V$-band luminosity for each $i$th galaxy, respectively. All nodes in the graph are connected with undirected edges. For each edge, we compute two values $\varepsilon_{ij} = (\alpha_{ij}, \varphi_{ij})$, which are the angles subtended by the central node to two other nodes (a toy illustration is shown in Figure 9), and the alignment angle between each node and the direction toward another node (a toy illustration is shown in Figure 10), respectively. The node and edge features are discussed in more detail in a subsection below and summarized in Table 2. Normalizing the 2D positions creates a GNN that is equivariant to permutations and invariant to rotations, reflections, and translations (see V. Garcia Satorras et al. 2021, for more details on equivariant GNNs). However, because one of our goals is to predict cluster orientation, our edge features are designed to provide the GNN with orientation information of the cluster substructure.

The core building block of our GNN branch is a layer that implements messages passing between connected nodes,

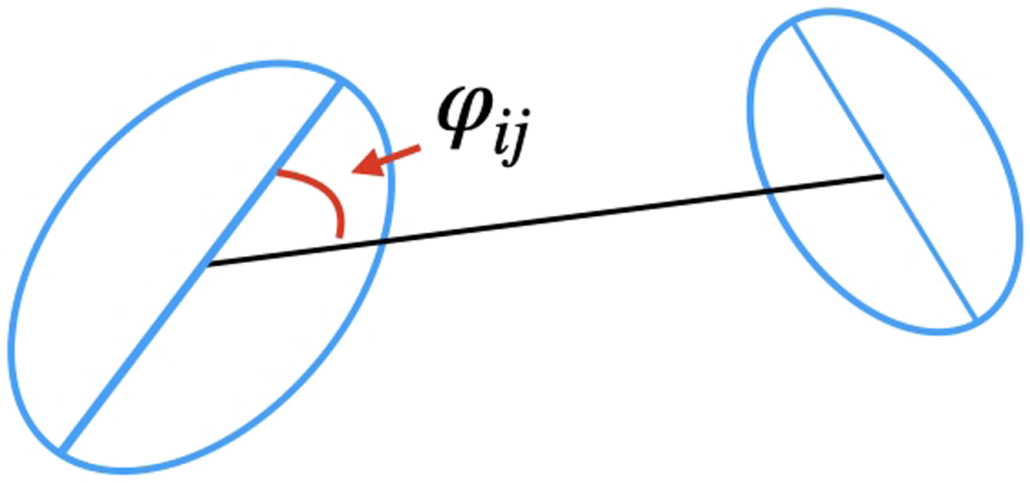

**Figure 10.** A toy illustration of edge feature angle $\varphi$ shown with two nodes representing a pair of galaxies. $\varphi$ is the alignment angle between the major axis of a galaxy and the vector connecting its pair.

**Table 2**
Descriptions of Input Training Features for Our hNN Network

| **Input Features** | |
|---|---|
| GNN Branch | |
| Node: $R_{\rm xy}$ | Relative 2D projected positions across our field of view w.r.t. the central node |
| Node: $v_{\rm rad}$ | Line-of-sight radial velocity |
| Node: $L_{\rm V}$ | $V$-band luminosity |
| Edge: $\alpha$ | Angle subtended by the central node and two other nodes |
| Edge: $\varphi$ | Alignment angle between each node the direction toward another node |
| CNN Branch | |
| Xray (soft) | X-ray photon intensity in range 0.5–1.2 keV |
| Xray (medium) | X-ray photon intensity in range 1.2–2.0 keV |
| Xray (hard) | X-ray photon intensity in range 2.0–7.0 keV |
| tSZ | Dimensionless $y_{\rm tSZ}$ parameter (Equation (8)) |

**Note.** The top section describes the input features of the GNN branch, where each feature is further labeled as a node feature or an edge feature. All input features of the CNN branch are normalized 128 × 128 images. Details of these features are described in Section 2.2.1.

similar to J. F. Wu & C. Kragh Jespersen (2023), where messages are constructed by concatenating the features of connected node pairs (source node $x_i$, receiving node $x_j$) with their corresponding edge attributes ($\varepsilon_{ij}$). This concatenated representation is processed through a three-layer MLP with layer normalization, GELU activations, eight hidden channels, and 16 latent channels.

The complete GNN branch consists of two sequential layers, each containing four unshared message-passing layers operating in parallel. The first sequential layer processes the original node and edge features, while each subsequent layer incorporates residual connections by adding the previous layer's output to the current layer's computation. This residual design helps preserve information flow through deeper networks and stabilizes training.

To generate graph-level representations from node-level features, we apply global mean pooling across all nodes within each graph, aggregating the final node embeddings into a single vector representation. This pooled representation serves as the graph-level feature vector that is subsequently passed to the feature fusion layer.

### 2.3.4. Task-specific Multilayer Perceptrons

Following feature extraction, the outputs from both the CNN and GNN branches are concatenated in the feature fusion layer, where they are passed through one layer of fully connected neurons. The architecture includes optional blend weighting to control the relative contribution of both CNN and GNN features before concatenation, though we did not find significant improvement using control blending, and our final architecture sets this to unity for equal weighting of both modalities. The fused representation is subsequently processed through an initial MLP, which performs regression to estimate triaxial lengths, after which the initial MLP output is concatenated with the original fused features (forming a skip connection) and passed to two parallel task-specific MLPs for regression of global shape variables and classification of the principal axis orientation variable. (The input and output variables are described in the subsections below.)

The regression MLP consists of three fully connected layers with progressive dimensionality reduction. Each layer incorporates layer normalization and GELU activation for training stability and nonlinearity. The classification MLP follows a similar three-layer structure but maintains consistent dimensionality throughout the hidden layers to preserve representational capacity for the multiclass prediction task. Each layer of this MLP employs GELU activation.

The dedicated output layers map the processed features to their respective targets: six continuous regression variables and one three-class classification variable.

### 2.3.5. Input Variables

*GNN node features:*

1. $R_{\rm xy}$, the 2D euclidean position in the plane of the sky for each galaxy.
2. $v_{\rm rad}$, the LoS radial velocity of each galaxy in a cluster given by:

$$v_{\rm rad} = [\boldsymbol{v}_p + H(z)a(z)\boldsymbol{r}_c] \cdot \frac{\boldsymbol{r}_c}{r_c}, \tag{10}$$

where $\boldsymbol{v}_p$ is the galaxy peculiar velocity given by MTNG, $H(z)$ is the Hubble parameter at redshift $z$, $a(z)$ is the scale factor at $z$, $\boldsymbol{r}_c$ is the comoving position to the cluster center, and $r_c$ is its magnitude.

1. $L_{\rm v}$, the $V$-band luminosity of each galaxy

*GNN edge features:*

1. $\alpha$, the angle subtended at the central galaxy by vectors to two other galaxies. A toy illustration is shown in Figure 9.
2. $\varphi$, the alignment angle between the 2D projected major axis (as given by the sorted eigenvectors of the second-moment tensor of each galaxy (Equation (1)) and the direction toward another galaxy. An illustration is shown in Figure 10.

We note that in order to obtain a reliable measure of angle $\varphi$, we restricted our analysis to galaxies resolved with at least 300 stellar particles. Galaxies with fewer than 300 stellar particles were treated as circular (as per Equation (6)) and assigned $\varphi_{ij} = 0$. We also experimented with including the measured $\epsilon_{\rm FoV}$ (Equation (6)) for each galaxy as a node feature, but found that this did not improve our model's performance.

**Table 3**
Descriptions of Output Features (Often Referred to as "Targets") for Our hNN Network

| | |
|---|---|
| **Output Variables (Targets)** | |
| Regression Tasks | |
| $\epsilon_+$ | 2D projected alignment along the $x$-axis of our FoV |
| $\epsilon_\times$ | 2D projected alignment along the $y$-axis of our FoV |
| $\Delta_\epsilon$ | difference between the minor to major axis ratio and the 2D projected ellipticity |
| $\lambda_a$ | length of major axis |
| $\lambda_b$ | length of intermediate axis |
| $\lambda_c$ | length of minor axis |
| Classification Task | |
| $\|e_{1_z}\|$ | orientation of principal axis (FoV, partial, or LoS) |

**Note.** All outputs are global properties of the cluster. The top section describes the output features assigned to the regression tasks, the bottom section summarizes the output features assigned to the multiclass classification task. Details of the input and output features are described in Section 2.2.1.

*CNN image features:*
The CNN branch is trained on four-channel 128 x 128 idealized images:

1. three channels are of X-ray photon intensity in soft (0.5–1.2) keV, medium (1.2–2.0) keV, and hard (2.0–7.0) keV
2. one channel is of $y_{\rm tSZ}$ effect (Equation (8)).

The details of the image construction are described in Section 2.2.2, and examples are visualized in Figure 6.

### 2.3.6. Output Variables

All of our output variables describe the global shape and orientation of our clusters.

1. $\epsilon_+$ and $\epsilon_\times$ (Equations (4)) describe the projected 2D alignment of the clusters along the $x$-axis and $y$-axis, respectively, across our FoV, and are used a sanity check to assess the quality of the hNNŠs predictions against quantities that can be measured in observations directly.
2. $\Delta_\epsilon$, the minor-to-major axis ratio corrected by the projected 2D ellipticity (Equation (7))
3. $\lambda_a$, $\lambda_b$, $\lambda_c$ the ordered axes lengths of each cluster, where $\lambda_a > \lambda_b > \lambda_c$
4. $|e_{1_z}|$, describes the orientation of the cluster principal axis as either FoV (across our FoV), partial (partially aligned between the FoV and Los), or LoS (oriented along our LoS).

Each output is described in detail in Section 2.2.1 and summarized in Table 3.

### 2.3.7. Training Procedure

We perform 10-fold cross-validation, each with an 80/10/10 split by partitioning the dataset into 10 equal subsets and training on eight while validating on one and testing on one, rotating through all combinations. This approach provides robust performance estimation and enables calculation of variance estimates across the different folds. Across the 10 folds, the standard deviation of the scoring metrics was $\lesssim$4% for six out of the seven parameters. The one exception was the parameter for $\lambda_a$, which has the greatest spread in its underlying data (spanning ∼0 to ∼10), which yielded a standard deviation of 5.3% in its metric. We selected the split with the best scoring metrics and performed a hyperparameter search in order to define a "fiducial" model. Based on a simple hyperparameter search, our hNN was trained utilizing an `Adam` optimizer (D. P. Kingma & J. Ba 2014), initially set with a learning rate of $8 \times 10^{-3}$ and weight decay of $10^{-4}$. Our batch size was 128.

We implemented a custom mixed loss function that combines both regression and classification objectives. The regression component uses the mean squared error (MSE) loss between continuous predicted and target variables, while the classification component employs cross-entropy for the multiclass prediction task. The mean of the combined losses (MSE loss for regression, cross-entropy loss for classification) is used to optimize the model.

A custom learning rate scheduling strategy is implemented to maintain training stability. The scheduler reduces the learning rate when the loss exhibits erratic oscillations, indicating that the model is stuck in a local minimum or experiencing gradient instability. However, if the learning rate drops below $10^{-5}$, it is automatically reset to the initial value to prevent premature convergence and allow continued exploration of the parameter space. During training, we inspect the training and validation losses to ensure that the optimization is converged. We implemented early stopping once the loss no longer decreased after 70 consecutive epochs, resulting in convergence after ∼300 epochs.

Lastly, to quantify algorithmic stochasticity due to random initialization of our fiducial model, we trained five independent models with the same data split and training procedure. Five runs allows for a reasonable balance between computational cost and confirming consistent model performance. We use the minimum and maximum values across the five runs to indicate uncertainty when reporting our model performance. The small standard deviation in the performance metrics across all five runs (<2% for most tasks and 4.7% for the most challenging task) indicates that additional training runs would not substantially change our conclusions.

### 2.3.8. Robustness Tests

We explored several changes to the architecture, hyperparameters, and training procedure; however, such tuning did not significantly affect the model performance. Some changes to the architecture included: making the network deeper/shallower by increasing/decreasing the number of layers in both CNN and GNN branches, increasing/decreasing the number of neurons in each dense layer, and trying different activation functions (ReLU and Leaky ReLU). We tuned hyperparameters by changing the values for the initial learning rate, and weight decay, and changing the scheduler and scheduler patience. We also tuned training by exploring image augmentation (randomly flipping and rotating the images) and found that image augmentation denigrated the model's performance in estimating $\epsilon_+$, $\epsilon_\times$, and $\Delta_\epsilon$. It is interesting to note, however, that the image augmentation did not significantly affect the model's performance with regards to predicting the other four output variables.

We also explored variations of input features, such as providing the Euclidean distance between two nodes as an edge feature in lieu of providing the 2D projected positions, $R_{xy}$, as node features in the GNN. We additionally tried providing the GNN with graph-level features pertaining to the brightest central galaxy (BCG). Studies have shown correlation, or mild misalignment (∼20°), between the major axes of the BCG and its host dark matter halo (R. Herbonnet et al. 2022; A. M. Delgado et al. 2023). Therefore, in an attempt to improve classification of principal axis orientation, we provided the GNN with (1) the orientation angle of the major axis of the BCG against the $x$-axis of the plane in the sky and (2) the ellipticity of the BCG, defined by $\epsilon_{\rm FoV}$, as two graph-level inputs in addition to the node and edge features described in Table 2. These two graph-level features pertaining to the cluster's BCG had no significant effect on model performance, so we did not include them in our final model.

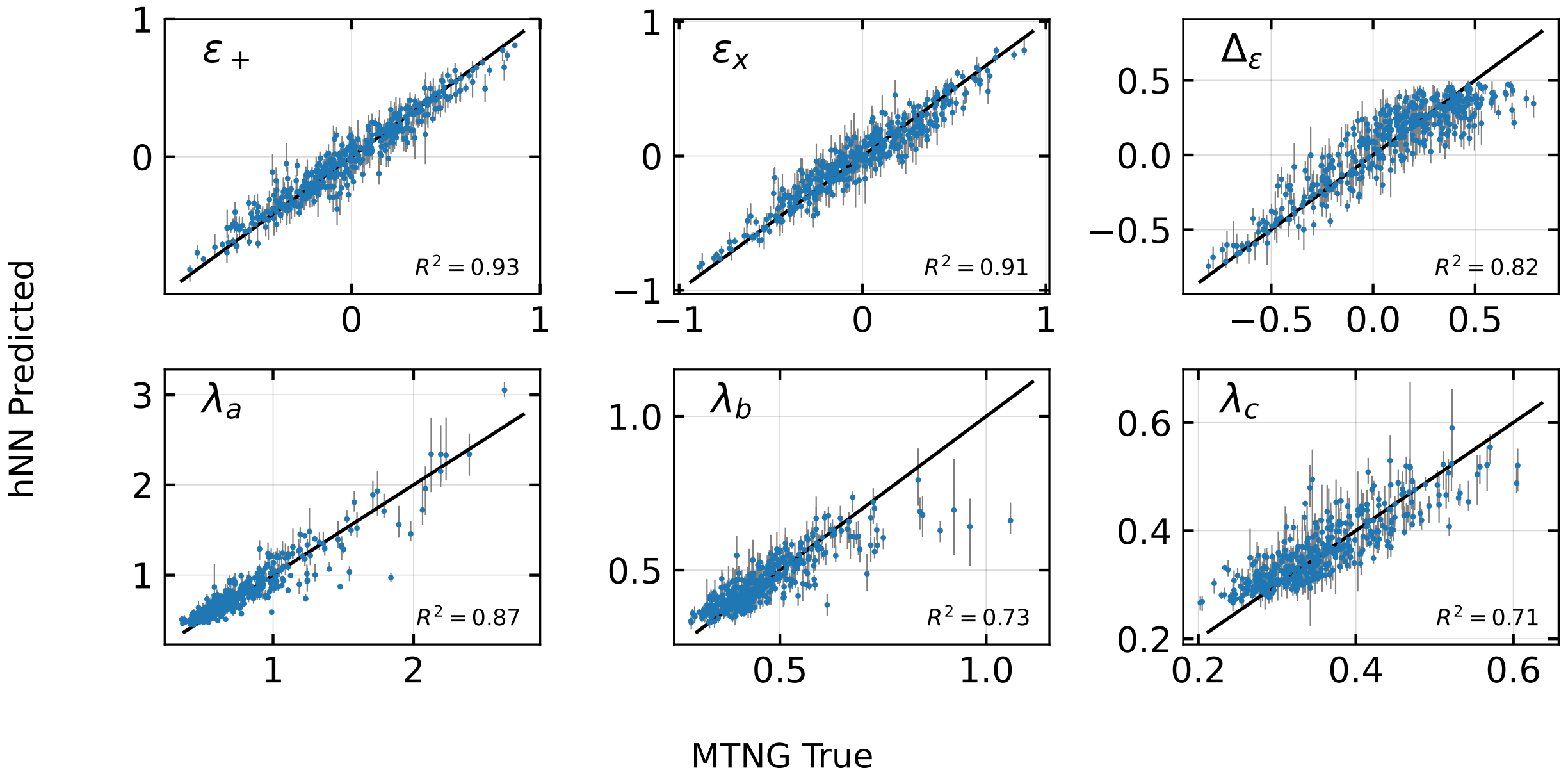

**Figure 11.** hNN regression results. Blue points show the mean prediction of five separate training runs of the hNN, and error bars (gray) are the upper and lower bounds of those runs. The black solid line depicts where a perfect prediction would fall. $R^2$ scores for each parameter are shown in the bottom-right corner of the panels. Our model is able to account for most of the variation in the data across all six variables. The axes of the top three panels (corresponding to variables $\epsilon_+$, $\epsilon_x$, and $\Delta_\epsilon$) are dimensionless, while those for the bottom three ($\lambda_a$, $\lambda_b$, $\lambda_c$) are in units of $h^{-1}$ Mpc. These results reveal the limitation of our model to distinguish between similar length axes in clusters: the bias in the prediction of $\lambda_b$ (the intermediate axis) levels off at the upper limit of $\lambda_c$ (the minor axis), and the bias in $\Delta_\epsilon$ levels off in the regime of spheres (where there is little to no difference in axis lengths.)

## 3. Results

In this section, we present the results from our novel hNN. Trained on 2D observables, our hNN estimates seven total variables including the 3D geometry of massive galaxy clusters. Six variables are estimated as regression tasks: the ordered axis lengths of the clusters (from longest to shortest—$\lambda_a$, $\lambda_b$, $\lambda_c$) and variables that describe global shape ($\epsilon_+$, $\epsilon_\times$, and $\Delta_\epsilon$). The seventh variable describes the principal axis orientation of clusters, $|e_{1_z}|$, and is a multiclass classification task (FoV, partial, LoS).

### 3.1. Regression Results: Cluster Triaxiality and Global Shape

Figure 11 shows results for regression tasks of the hNN. The top three panels show results for the dimensionless "global shape" features of the clusters, and $\Delta_\epsilon$ (Equation (7)). The bottom three panels show results for the lengths of the ordered triaxial axes, $\lambda_a$, $\lambda_b$ and $\lambda_c$ in units of $h^{-1}$ Mpc. Blue data points are the mean prediction from five independent training runs of the hNN. Error bars (in gray) indicate the range (minimum and maximum values) across the five runs. Five runs is sufficient for reporting the uncertainty. For model evaluation, we use the $R^2$ score for the mean prediction across all five runs, defined as follows:

$$R^2 = 1 - \frac{\sum_{i=1}^{n}(y_i - \hat{y}_i)^2}{\sum_{i=1}^{n}(y_i - \bar{y})^2}, \tag{11}$$

where $y_i$ is the observed value as per MTNG, $\hat{y}_i$ is the mean predicted value for sample $i$ across all five runs, and $\bar{y}$ is the mean of the observed values. The $R^2$ score, commonly referred to as the "coefficient of determination" or the "goodness of fit," measures the proportion of variance of the target variable explained by the model.

Unsurprisingly, hNN performed very well in predicting 2D projected ellipticities, $\epsilon_+$ and $\epsilon_x$ (Equations (3) and (4)), with $R^2 \sim 0.90$, indicating that 90% of the data variance is explained by our model for these variables. These variables can be directly measured from the positions of the galaxies in a cluster across the FoV. Given that the graph is rotationally invariant, we use these two variables as a sort of "sanity check" to confirm that the network can accurately predict features that we can measure directly.

The hNN was able to account for ∼81% of the variation of $\Delta_\epsilon$ (Equation (7)). The top-right panel of Figure 11 shows a slight bias toward the mean, with values $\Delta_\epsilon > 0.5$ consistently being underpredicted. While the poorer estimation in this regime could be attributed to the small sample size (there are only 29 clusters with $\Delta_\epsilon \gtrsim 0.5$ in the test set), these 29 data points have a mean absolute percentage error (MAPE) of 0.323 from truth. Comparably, there is a small sample size at the opposite end of the $\Delta_\epsilon$ distribution (22 clusters with $\Delta_\epsilon \lesssim 0.5$ in the test set); however, these 22 data points have an MAPE of 0.111 from truth. We remind the reader that as $\Delta_\epsilon \to 1.0$, clusters become increasingly spherical, where there would be no meaningful distinction between the minor and major axes of the cluster, although their ratio is necessary in the estimation of $\Delta_\epsilon$.

An $R^2 = 0.85$, as seen in our regression result in $\lambda_a$, suggests that 85% of the data variance is explained by our model. The scores measured for $\lambda_b$ and $\lambda_c$ are $R^2 = 0.70$ and $R^2 = 0.65$, respectively. In most clusters, the intermediate and minor axes lengths are very similar, and the hNN may struggle to distinguish between the two. This is further evidenced in the bias for $\lambda_{\rm b}$, which levels off at a value comparable to the upper limit of true $\lambda_{\rm c}$ values.

### 3.2. *Classification Results: Cluster Orientation*

Determining the orientation of massive clusters in addition to the triaxial axis lengths is of particular importance because errors on cluster mass estimates arise from the oversimplification of treating prolate clusters as spherical, particularly when these clusters are oriented along our LoS.

Principal axis orientation is a particularly challenging problem due to the fact that most clusters in our catalog are either spheroidal or have ambiguous morphology; thus, they have ambiguous orientation. Our approach was to treat this as a classification problem since we care most about identifying which clusters are aligned along our LoS. We evaluate the model using the built-in accuracy score function from `Scikit-Learn`,[10] which for multiclass scoring returns the subset accuracy defines as follows:

$$\text{Accuracy} = \frac{1}{n}\sum_{i=1}^{n}\mathbf{1}(y_i = \hat{y}_i), \tag{12}$$

where $n$ is the total number clusters in the test set, $y_i$ is the true label of the $i$th cluster as per MTNG, $\hat{y}_i$ is the hNN predicted label, and 1 is the indicator function (1 if the condition is true, 0 otherwise).

Figure 12 shows the hNN classification of the principal axis orientation, $|e_{1_z}|$, resulting from our fiducial model. The top panel shows the confusion matrix results from the full test set for $|e_{1_z}|$ of all clusters. We report a ∼58% accuracy averaged for all classes for our fiducial model. The highest frequency corresponds to the correct classification, 68%, of clusters elongated along our LoS. Errors are the minimum and maximum values from five independent training runs of the hNN. The bottom panel is similar to the top panel but only for the subset of false spheres found within the full test set. The hNN was able to accurately classify ∼71% of these rare instances as a subset of classified LoS-aligned clusters.

Figure 13 illustrates a method for evaluating our model's ability to accurately estimate the triaxiality of a cluster conditioned on correct orientation classification. For clusters with correctly classified $|e_{1_z}|$, we then measure the percentage of clusters where all three axis lengths, $\lambda_{\rm a}$, $\lambda_{\rm b}$, $\lambda_{\rm c}$, are estimated within an absolute error threshold of MTNG defined as follows:

$$\text{Percent Clusters} = \frac{1}{N}\sum_{i=1}^{N}\begin{cases}1 & |y_{i,x} - \hat{y}_{i,x}| < \tau, \\ & \text{for all} x \in \{a, b, c\}, \\ 0 & \text{otherwise}\end{cases} \tag{13}$$

[10] https://scikit-learn.org/

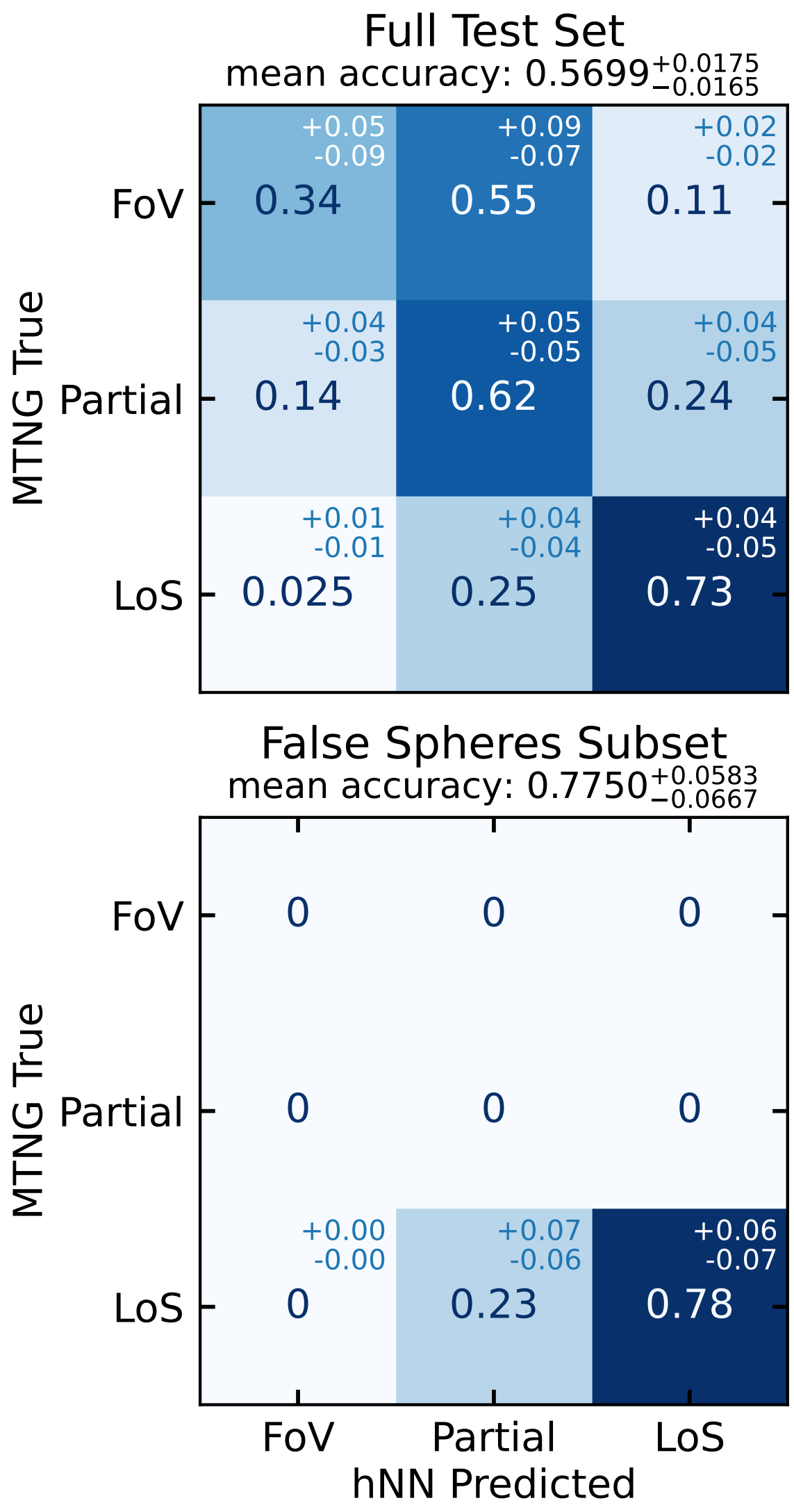

**Figure 12.** Mean classification results from our five independent training runs of the hNN. Top panel: shown here is the confusion matrix results from the full test set for the classification of the principal axis orientation, $|e_{1_z}|$, of all clusters. The accuracy averaged for all classes is reported above the confusion matrix. The highest frequency, with a mean of 0.73, corresponds to the correct classification of clusters elongated along our LoS. Errors are the upper and lower bounds calculated from the five separate training runs. Bottom panel: the image here is similar to that of the top panel but only for the subset of false spheres found within the full test set. The hNN was able to accurately classify a mean of 0.78 of the false spheres in the test set.

where $N$ is the total number of clusters in the sample, $y_i$ is the $i$th observed value as per MTNG, $\hat{y}_i$ is the corresponding hNN predicted value, $x$ indicates the axis, $a$, $b$, or $c$, and $\tau$ is the percent threshold. The $y$-axis of Figure 13 is the percent of clusters in the test set, and the $x$-axis is the percent threshold, $\tau$. Our fiducial model is shown by the dashed black line. We compare our fiducial to two hypothetical models that guess spherical shapes: one conditioned on predicting the correct orientation but guessing the spherical mean of MTNG for all three axis lengths (blue dashed–dotted line) and another that randomly guesses orientation and guesses the spherical mean of MTNG for the axis lengths (blue dotted line). The top black dotted line shows the maximum possible percentage based on the conditioning on correct orientation, which is ∼58%. We see that our fiducial model outperforms a spherical model by ∼20%–30% for absolute error thresholds <0.20 of MTNG.

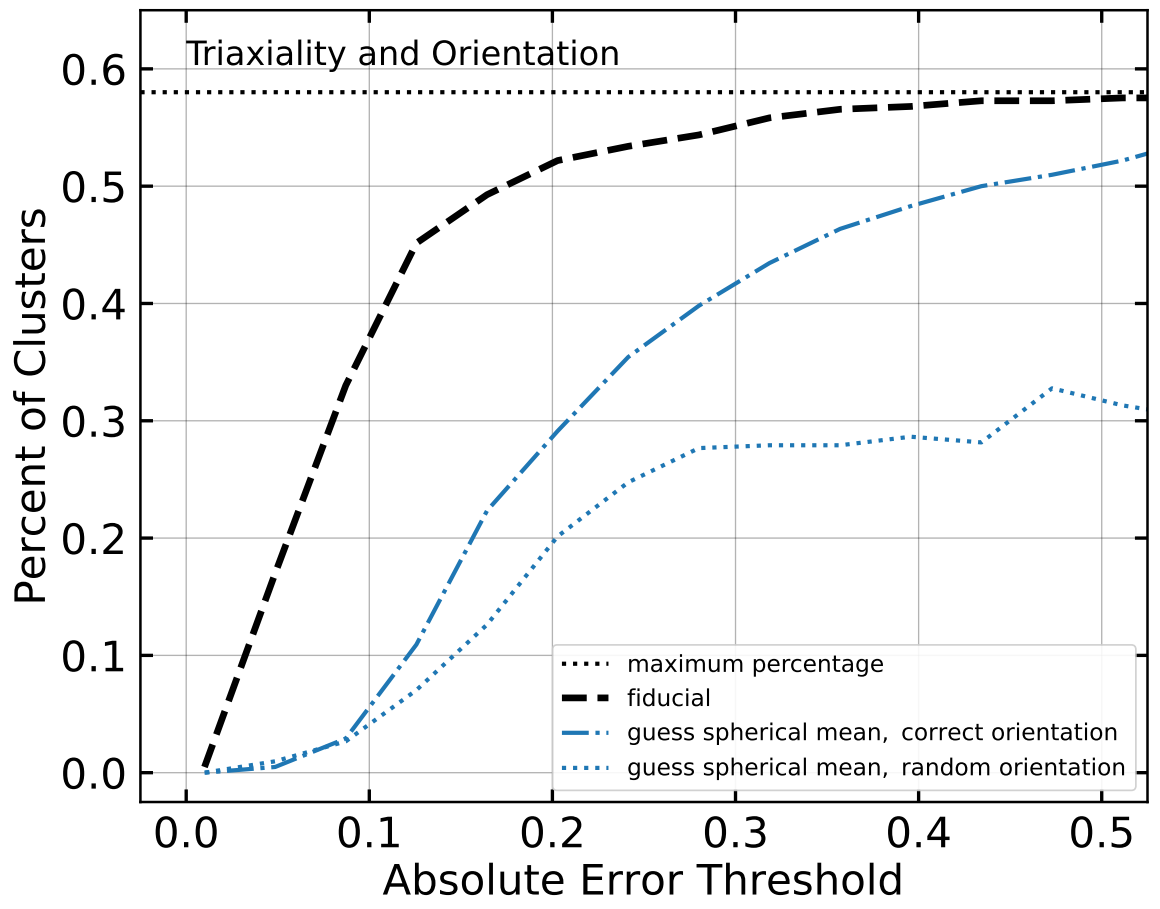

**Figure 13.** The percent of clusters in the test set for which orientation was correctly classified and triaxiality (all three semiaxes) was estimated within an absolute error threshold of MTNG True. We only show the absolute error within 50% of MTNG (along the *x*-axis). The dashed black line is the result from our fiducial hNN model. We compare our model against two hypothetical models: one predicting correct orientation but guessing the spherical mean of MTNG True in the dotted–dashed blue line and another that guesses the spherical mean and gives a random orientation for the cluster in the dotted blue line. We see that our fiducial model outperforms guessing a spherical shape in either case. The dotted black line at the top is the maximum percentage possible given the condition of correct orientation classification.

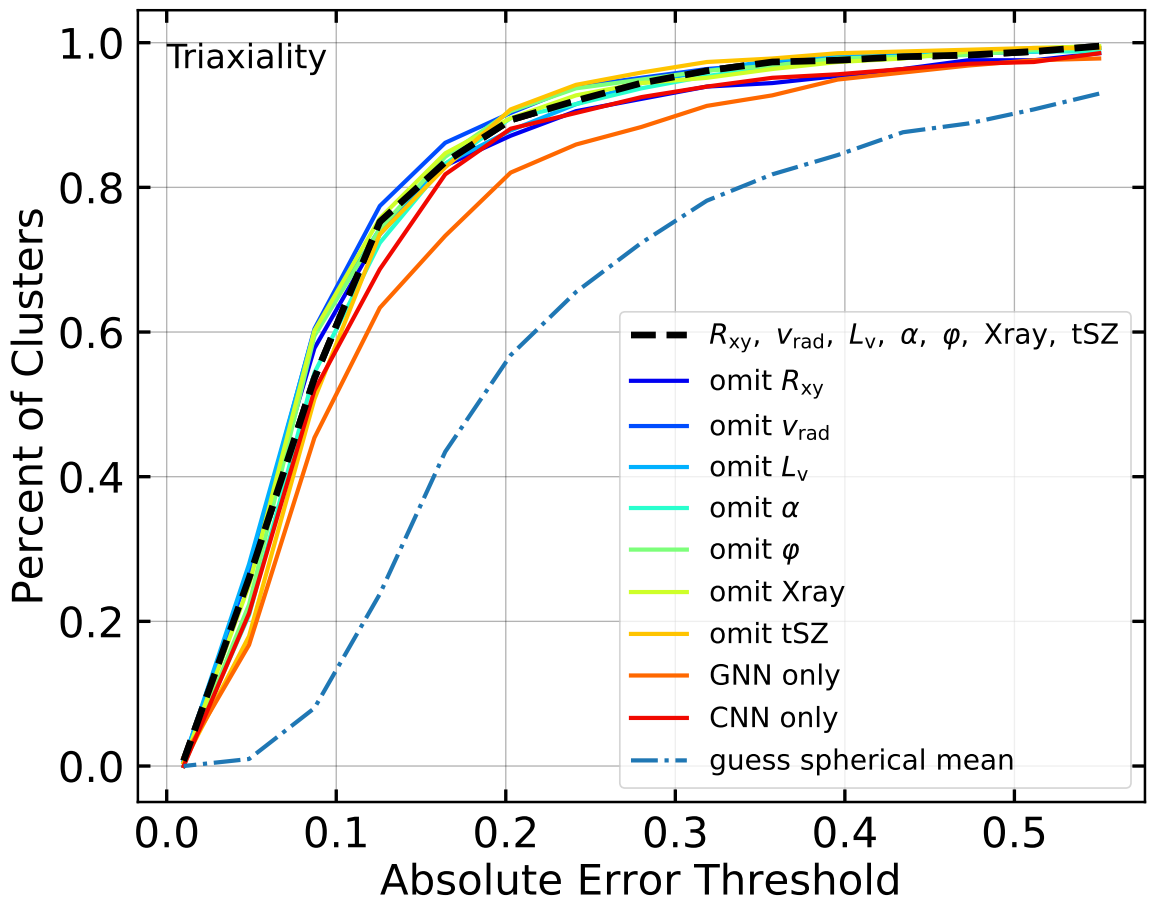

**Figure 14.** An approach to feature importance analysis. We performed an ablation study by systematically omitting features or channels and retraining models with identical architecture, hyperparameters, and duration as our fiducial model. The above results show the percentage of test set clusters within a given threshold of MTNG (similar to Figure 13) for triaxiality regression only (without conditioning on classification). Our fiducial model (dashed black line) is comparable to ablated variants, with the four-channel image-omitted, GNN-only model showing the largest performance degradation, indicating the importance of the CNN branch in our model. All variants are within ∼15% of each other and significantly outperform a hypothetical model that guesses the spherical mean of MTNG True (dotted–dashed blue line).

### 3.3. Model Analyses

Model interpretability is particularly challenging in a deep learning approach where the output is not an equation. In order to analyze which components of our model are important, we performed an ablation study. An ablation study is performed by systematically omitting training features or channels (typically done one at a time) and retraining the models with identical architecture, hyperparameters, and duration. Our ablation study consisted of 31 models with various combinations of the training features and channels included in our fiducial model. (The 31 combinations can be seen in the legend of Figure 16.)

Figure 14 shows results as per Equation (13) for only triaxiality. It is similar to what is shown in Figure 13 but not conditioned on classification, so that we can analyze the effects on triaxiality estimation alone. All models performed within 15% of fiducial. The models in Figure 14 are those that only omit one feature, branch, or channel. We see that the GNN-only model has the most degradation in performance, indicating the importance of the four-channel CNN for estimating triaxiality. The CNN-only model performs comparably to the various hybrid models, and the graph features improve performance on triaxiality only by ∼5%.

Classification results for the ablation sample in Figure 14 are shown in Figure 15. All variants have an average accuracy within 10% of our fiducial model; however, we note there are several models for which the frequency of the LoS clusters is higher than our fiducial. In general, all models struggle to classify the orientation of FoV clusters, which is most likely due to the noise created by spheroids, whose symmetry makes determining principal axis orientation a challenge for this particular task. We also note that the model that omits $\varphi$ (center panel) has the highest accuracy. $\varphi$ provides the alignment angle between a galaxy and the vector to another galaxy. Galaxies tend to be intrinsically aligned along the underlying distribution of matter (F. Valdes et al. 1983; J. Miralda-Escude 1991; C. M. Hirata et al. 2007), and we therefore expect there to be important group orientation information encoded in the alignment of galaxy members. We remind the reader, however, that we only confidently measured the alignment angle for galaxies resolved with at least 300 stellar particles. For all other galaxies (resolved with only minimum of 50 stellar particles), we assumed a circular shape with an alignment along an imaginary *x*-axis. Therefore, each cluster only has alignment information for a limited number of massive galaxies. The models with the two most degraded performances are the GNN-only and CNN-only models, indicating the importance of the hybrid model for extracting information from clusters across multiple probes.

Figure 16 shows the percent of clusters within a threshold of MTNG conditioned on correct classification (similar to Figure 13) for all 31 models in our ablation study. The legend is ordered by model performance at the 0.10 absolute error threshold, where the top model is the highest performing. It shows that the model which omits $\varphi$ (top model) not only has the highest maximum percentage (as per the classification condition) but is also the highest-performing model at the 0.10 threshold. We would therefore recommend omitting $\varphi$ as a feature unless we can confidently measure this value for all of the galaxy members in a cluster.

#### 3.3.1. Image Analysis

Interpreting deep learning methods is a challenging task due to their highly nonlinear nature. One approach to interpreting CNNs is to determine what the network is "paying attention to" in an image. In order to further interpret the feature extraction of our hNN, we employ saliency maps as a means for interpreting the CNN branch of our fiducial model. Many deep-learning-based X-ray and SZ cluster mass studies use some version of saliency, or pixel activation, to analyze cluster regions deemed significant by their CNNs (M. Ntampaka et al. 2019; Z. Yan et al. 2020; D. de Andres et al. 2022; M. Ho et al. 2023). We utilize standard saliency maps, by which we obtain the absolute values of the gradients of our outputs w.r.t. each pixel and produce heatmaps of the activated pixels. Each pixel of our $128 \times 128$ images corresponds to $\sim$62.5 kpc in spatial length, and the pixels are normalized to receive a value between 0 and 1, where 0 is nonactivated.

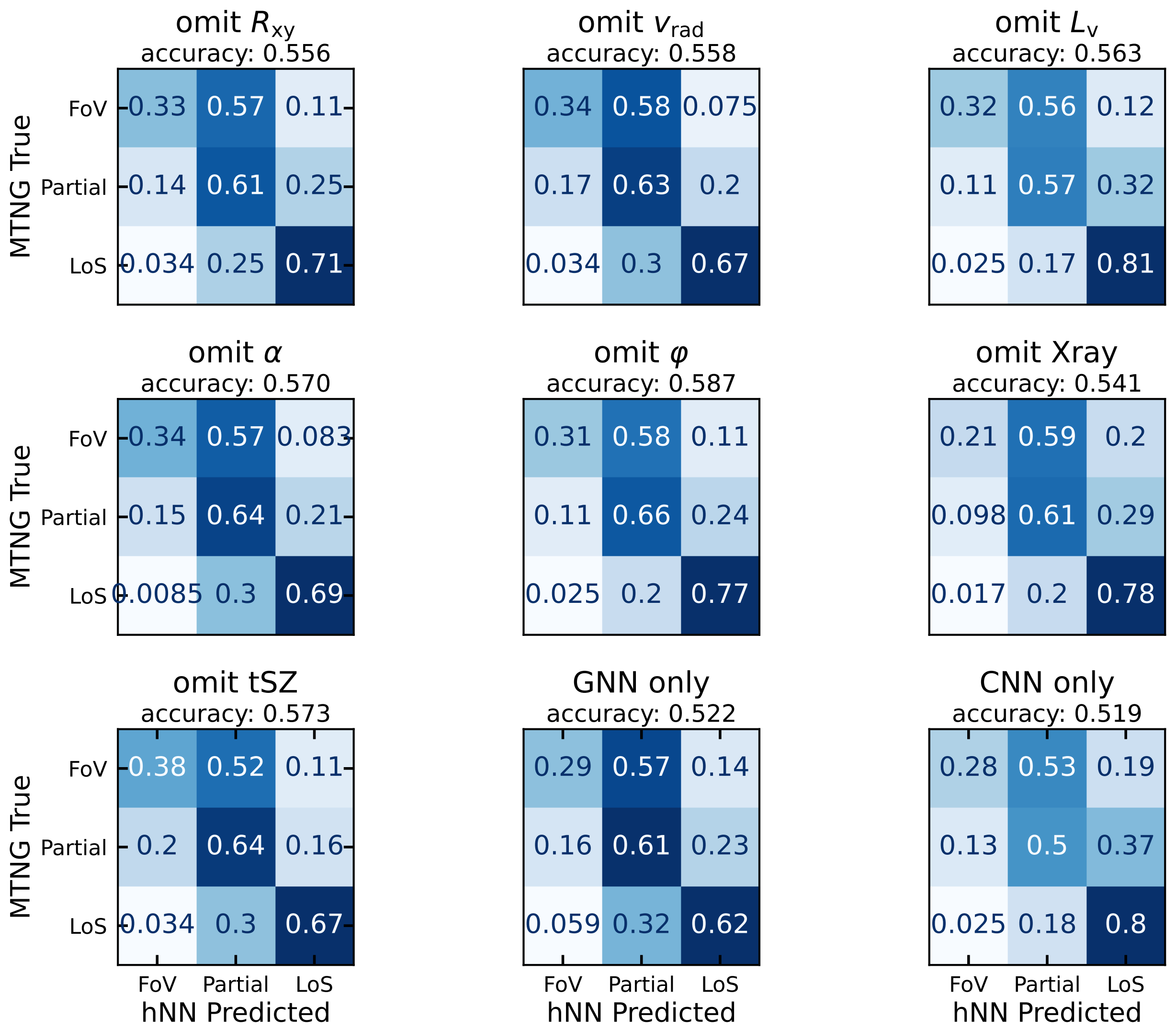

**Figure 15.** Feature importance analysis via ablation study for the classification task. Confusion matrices are shown for each of the ablated variants. Accuracy scores are also reported for each model. Results from our fiducial model (top panel of Figure 12) are comparable to most of the variants. We emphasize that these are results from an idealized study, and we caution against over-interpretation of any quantitative performance metrics.

Figure 17 shows mean population saliency maps of the test set. The figure is organized with each band in our four-channel images (rows) as a function of cluster mass (columns). Each map corresponds to the mean pixel activation from stacked test images for the given band and mass bin. We note that there are a decreasing number of clusters within increasing mass bins, with 238, 120, 37, and 17 clusters in each bin to be precise. This explains why the maps in the highest-mass bin appear more diffuse compared to those in the lowest-mass bin. Interestingly, the CNN tends to ignore the cluster centers, in agreement with prior deep learning studies on X-ray cluster masses, which found that image pixels corresponding to ICM regions were significantly more activated than core/AGN regions (M. Ntampaka et al. 2019; M. Ho et al. 2023). The central-most regions of clusters are very bright in both the X-ray and tSZ, and it is likely that the immense density of cluster cores would not be distinguishable between clusters, thus not providing discernible information for the network. It is reasonable to interpret that the CNN would instead pay more attention to regions where there are nuanced differences between instances, such as in intermediate and outer regions, which are indeed where the most activated pixels lie. This also supports observational studies where they find that omitting cluster core regions results in reduced scatter in cluster observable–mass relations (B. J. Maughan 2007; A. B. Mantz et al. 2018).

We additionally examined saliency maps of individual clusters for specified output variables and show the results in the Appendix. Figures 18–23 are organized such that each row is an individual cluster, columns (1)–(4) (left to right) show saliency maps for one channel as labeled, and the fifth-rightmost column shows the consensus of activated pixels across the four multiwavelength channels. We define consensus as a pixel having a minimum 0.1 (out of a scale of 0–1) activation value in each channel. Notably, the consensus maps reveal that many clusters have little to no consensus across channels. This implies that the contribution from each wavelength is quite different, highlighting the importance of a multiwavelength approach to galaxy cluster studies.

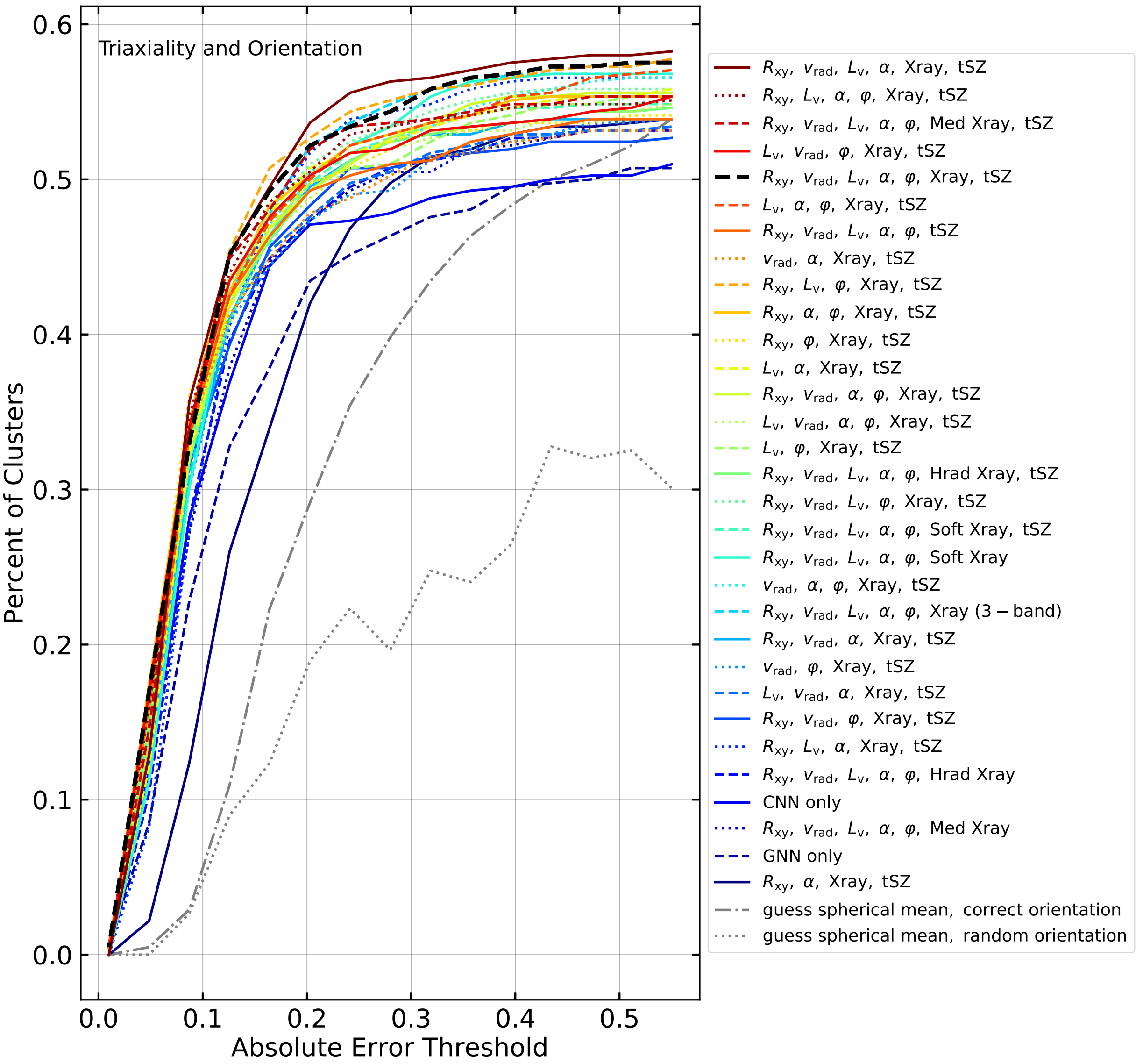

**Figure 16.** Similar to Figure 13 but here showing results for the full ablation study. The models are color coded by performance at the 0.10 absolute error threshold. The legend displays the models ordered from (top to bottom) best performing to worst performing. Our fiducial model is shown as the bold, dashed line in black.

## 4. Summary and Discussion

Massive galaxy clusters are powerful probes of cosmology, but their triaxiality and orientations are systematics that affect cluster selection and mass estimation methods. We have designed an hNN (often referred to as a fusion network) that takes idealized 2D observables of massive galaxy clusters from the flagship hydrodynamical volume of the MillenniumTNG (MTNG) project to predict 3D triaxiality and principal axis orientation. For this study, we consider only clusters with critical mass $M_{200} \geqslant 10^{14}\,M_{\odot}$ (total mass enclosed in a sphere whose mean density is 200 times the critical density of the Universe), resulting in a catalog of 4117 massive clusters.

We began this study by investigating cluster morphology in MTNG. We defined several variables for measuring both projected and 3D cluster morphology, described in Section 2. Although roughly one-third of the clusters in our catalog have an ambiguous or irregular shape, we were able to define three clear morphologies for the other two-thirds of the catalog: elongated ellipsoids, false spheres, and spheroids. We have a particular interest in identifying false spheres, which are rare clusters that appear to be spherical across our plane in the sky due to their projected circular shape, but are, in fact, prolate ellipsoids elongated along our LoS. False spheres in particular can lead to biased estimates in weak lensing studies and richness studies if assumed to have a spherically symmetric geometry. We identified these rare objects as those with $|\Delta_{\epsilon}| < 0.2$ and $|e_{1_z}| > 0.75$ (described in Table 1).

Designing a method to classify false spheres directly from 2D observables would be particularly challenging given their rare frequency (there are only 380 total in our catalog and only 24 in our test set). Instead, we designed an ML approach to estimate the triaxial lengths and principal axis orientation of all clusters, this way "capturing" the false spheres within the sample.

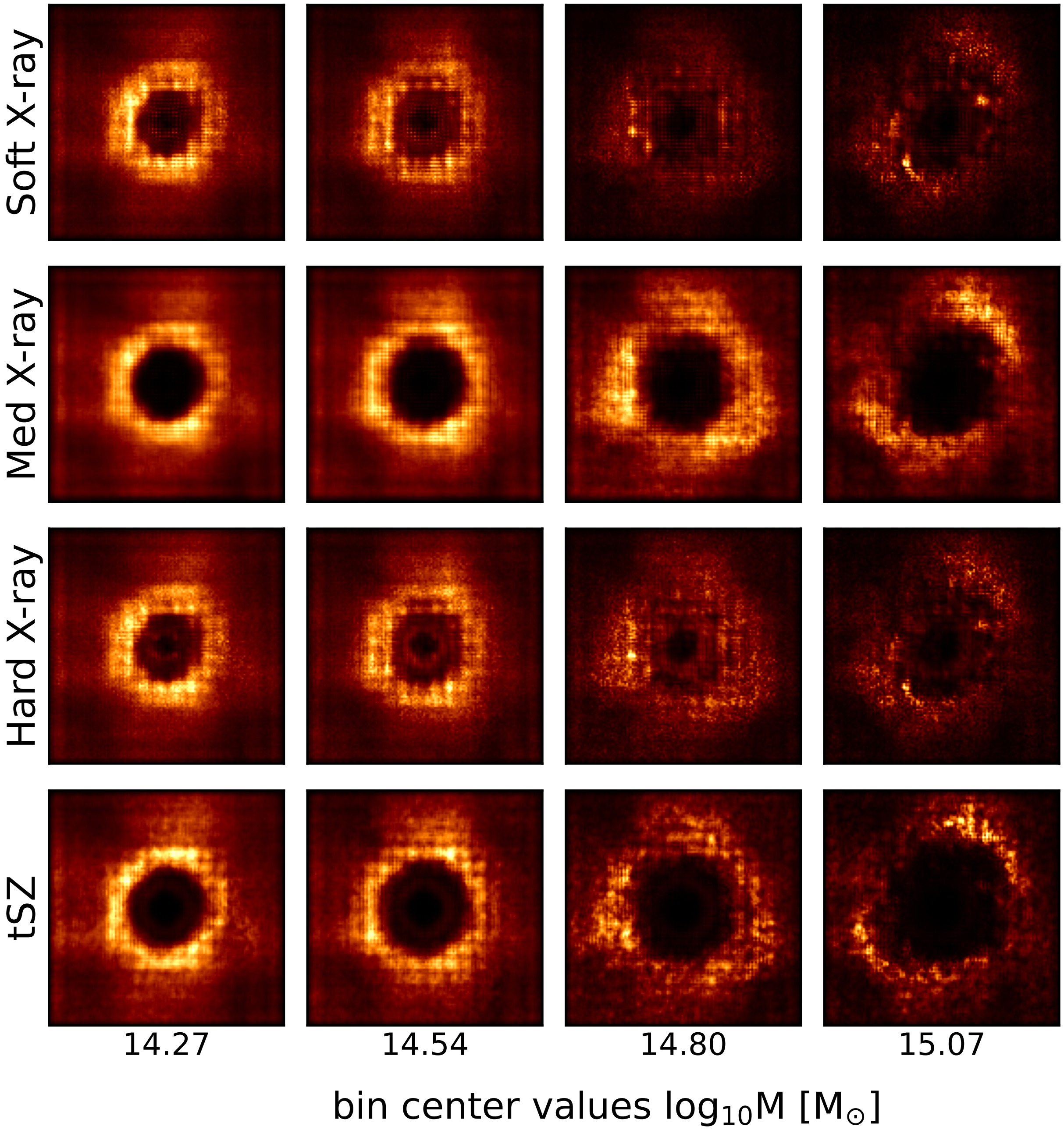

**Figure 17.** Mean population saliency maps from the CNN branch of the network show by wavelength (rows) as a function of binned mass (columns). We note that the number of clusters in each column decreases as mass increases, with the lowest-mass bin containing 238 clusters, while the highest-mass bin contains 17 clusters. Each map shows the mean of activated pixels from the stacked test set. The centers of the galaxies are largely ignored by the network. However, the soft X-ray and hard X-ray maps do show activation of pixels closer to the center than the medium X-ray and tSZ maps.

Our hNN architecture consisted of two input branches: a GNN branch for cluster data similar to those available in optical or NIR surveys (2D projected galaxy positions, galaxy LoS radial velocities, and luminosity in the $V$ band), and a CNN branch for four-channel multiwavelength cluster images (soft X-ray, medium X-ray, hard X-ray, and tSZ). Each of the specialized branches perform feature transformation via message passing (GNN) and feature extraction (CNN). The latent representations from the respective branches are first sent to a fusion combination layer where they are further encoded through an initial dense layer of neurons before being sent off to three different task-specific MLP. One MLP performs regression for estimating the ordered triaxial lengths of the clusters (from longest to shortest—$\lambda_a$, $\lambda_b$, $\lambda_c$). This MLP then provides input via skip connection to the other two parallel MLPs, one that performs regression for estimating variables that describe the global shape of our clusters ($\epsilon_+$, $\epsilon_\times$, $\Delta_\epsilon$), and a third MLP that performs classification to predict the principal axis orientation (FoV, partial, LoS).

Finally, we perform model analysis via an ablation study in which we systematically omit training features or channels from our pipeline in order to determine their importance. We perform this for 31 different combinations and retrain the models with an identical architecture (where applicable), hyperparameters, and duration as the fiducial model. We further examine saliency maps for each wavelength in the four-channel images in order to analyze the feature extraction of the CNN branch.

We wish to re-emphasize the idealized nature of the simulated inputs in this study and the complexity of observational systematics in real-world surveys. The reported metrics in this study, therefore, reflect this model's performance under idealized conditions and are expected to decrease when applied to real observational data. Our main findings can be summarized as follows:

i. Our model is able to account for ∼87% of the mean variation in the major axis lengths, $\lambda_a$, of our test set clusters. However, the same model can only account for mean scores of ∼73%, 71% for $\lambda_b$ and $\lambda_c$, indicating that the hNN may struggle to distinguish between the intermediate and minor axes (Figure 11). Nevertheless, the hNN is able to estimate all three axis lengths to within 0.1 absolute error of MTNG for ∼60% of clusters in the test set. This is $\gtrsim$40% improvement compared to approximating a simplified, spherical shape, which would only result in <20% of clusters in the test set meeting the same threshold. For the small number of false spheres in the test set, we capture ∼42% at the threshold of 0.1 within MTNG (Figure 14).

ii. Our model is able to classify the principal axis orientation (FoV, partial, or LoS) with 57% accuracy on average, with the highest frequency, 73%, for the orientation of prolate clusters oriented along the LoS. It struggles to classify FoV orientation (oblate clusters whose principal axis is oriented across our FoV), preferentially classifying them as having a "partial" alignment. This is unsurprising since, with the exception of oblate elongated ellipsoids, determining the orientation of symmetric or irregular morphologies is challenging. The model is further able to accurately capture 78% of false spheres (prolate clusters elongated along our LoS). There is a slight overabundance of prolate spheroids in our cluster catalog, and further investigation is required to determine whether the false spheres are captured in the "net" of prolate spheres or of prolate ellipsoids (Figure 12).

iii. Overall, our fiducial model is able to estimate the triaxiality conditioned on correctly classified orientation of ∼30% more clusters to within 0.1 absolute error of MTNG compared to approximating a simplified, spherical shape (Figure 13).

iv. Results from our ablation study underscore the power of combining the GNN and CNN branches for cluster shape estimation. The results reveal that most ablated variants are within 10% of the fiducial, with the CNN-only and GNN-only models having the most performance degradation. Additionally, omitting $\varphi$ (the alignment angle of a galaxy member with the connecting vector to another galaxy member) resulted in only a marginal increase to the model performance, which likely stems from reliable measurements of this parameter being limited to massive galaxies only (Figures 14, 15, and 16).

v. Analysis of saliency maps (pixel activation) across the four-channel images from the CNN branch reveals discernible differences in contribution from each channel, highlighting the importance of a multiwavelength approach to cluster studies (Figures 18–23).

As discussed in Section 1, the morphology of clusters is highly influenced by their formation histories. Mass accretion rate, for example, has an important role in determining cluster properties such as radial density profile (B. Diemer & A. V. Kravtsov 2014), ellipticity, and mass fraction contained in the substructure (K.-W. Wong & J. E. Taylor 2012; F. Jiang & F. C. van den Bosch 2016). A strong estimator of cluster geometry can thus also aid in further understanding its formation history. A study by J. Soltis et al. (2025) utilized the same cluster catalog as here and designed a deep learning model which took the four-channel images, with a simulated Chandra instrument profile applied, as inputs in order to create a mass accretion rate (MAR) estimator. Their model was nearly a factor of 2 more precise than mass-based MAR estimates. They investigated the importance of substructure and ellipticity in their model by decomposing the cluster images into symmetric and asymmetric components. In one experiment, they passed only the symmetric components of the images to the MAR estimator and, as expected, they found that the model systematically underestimated the MAR of clusters when given only the symmetric components. It would thus be interesting to build on these ideas and design a geometry-based MAR estimator. If cluster triaxiality can be used as an MAR estimator, it would further underscore the importance of correctly modeling 3D cluster geometry.

Accurately estimating the 3D geometry of massive clusters from 2D observables by utilizing, or building upon, our novel hNN can help address various systematics in cluster cosmological studies (as described in our introduction). One example of the advantage of using our method is to obtain a concentration–mass relation for halos by fitting triaxial density profiles instead of spherically averaged ones, which will be done in our forthcoming study. Triaxial clusters aligned along the LoS analyzed under spherical assumptions can lead to artificially high concentration parameters (M. Oguri & R. D. Blandford 2009; M. Sereno et al. 2010; A. M. Groener et al. 2016). Because the concentration–mass relation is sensitive to cosmology, we expect that a more accurate measure of halo concentration, by way of halo triaxiality, will provide tighter cosmological constraints.

Recent works using fully triaxial forward models for galaxy clusters in the Cluster Heritage project with XMM-*Newton*–Mass Assembly and Thermodynamics at Endpoint of structure formation (CHEX-MATE; CHEX-MATE Collaboration et al. 2021) have underscored this point. H. Saxena et al. (2025) employed a Monte Carlo (MC) framework to fit triaxial Navarro–Frenk–White density profiles (J. F. Navarro et al. 1997) directly to mock SZ-selected clusters, where the selection follows the Planck SZ detection pipeline (Planck Collaboration et al. 2016b). In their analysis, the intrinsic axis ratios and viewing angles of the clusters are marginalized over. They show an orientation bias in the SZ selection, where the pipeline preferentially selects prolate clusters oriented along the LoS. This, in turn, produced biases in weak-lensing derived mass estimates when the clusters are modeled spherically. In this context, a fast ML-based predictor for intrinsic 3D shape and orientation, such as our hNN, can be used to inform the priors of the MC analysis or to provide a first-order correction for orientation bias in SZ samples.

Similarly, A. Gavidia et al. (2025) carried out a full triaxial analysis of galaxy clusters using joint SZ, X-ray surface brightness and temperature, and weak-lensing shear data. They adopted a forward-modeling MC approach in which a triaxial mass and gas distribution is projected into each observable. They fist validated the method on a mock CHEX-MATE cluster catalog and then applied the method to the real system A1689. They compared triaxial and spherical fits, and showed that allowing for triaxiality changes the inferred mass and concentration, as well as the nonthermal pressure fraction. They further showed that their results are in excellent agreement with prior triaxial and spherical studies of A1689. Our hNN approach is complimentary to these studies. Our network, trained on MTNG clusters, learns the nonlinear mapping from the 2D multiprobe observables directly to the 3D cluster geometry. This offers a path to propagating realistic triaxial shapes into large cluster samples, thereby addressing triaxiality-driven systematics in mass and concentration estimates in a statistically viable yet computationally efficient way.

Improved measurements of clusters that occupy the high-mass end of the halo-mass function can help put tighter constraints on Λ cold dark matter. Our novel approach offers a method for addressing the impact of triaxiality on cluster statistics, by leveraging deep learning techniques trained on state-of-the-art hydrodynamical simulations to extract nonlinear features from massive clusters. This represents an advancement over standard approaches that rely on simplified models, whereas deep learning can capture subtle, multidimensional relationships for more nuanced predictions that would be difficult to achieve with traditional methods.

## Acknowledgments

We wish to thank the anonymous referee for providing excellent feedback, which helped strengthen the quality of this paper. We also wish to thank Benjamin Wandelt, Niall Jeffrey, Matt Ho, Shivam Pandey, Nico Garavito Camargo, Adrian Bayer, and Max E. Lee for interesting and helpful discussions during this work. S.B. is supported by the UKRI Future Leaders Fellowship (grant Nos. MR/V023381/1 and UKRI2044). The material presented is based on work supported by the National Science Foundation (NSF) grant 2307070 CDS&E: A Modern Toolkit to Enhance the Science Productivity of Optical Survey Data. This work was carried out at the Advanced Research Computing at Hopkins (ARCH) core facility (rockfish.jhu.edu), which is supported by the NSF grant No. OAC 1920103.

## Data Availability

All simulation data of the MillenniumTNG project will be made publicly available on a dedicated website, currently planned for by the end of 2026. The MTNG galaxy cluster data used in this study will be shared upon email request to the corresponding author.

## Appendix
## Individual Saliency Maps

The saliency maps in this appendix (Figures 18–23) show activation for individual clusters and specific output variables. Each figure is titled with the output variable. Each row in each figure is a different cluster, whose morphology and true orientation are furthermore labeled in the leftmost saliency map. The first four columns show activation at each wavelength (as labeled), and the fifth column shows mean consensus activation maps, where each pixel in the consensus map is activated only if the pixel received $\geqslant$0.10 activation at every wavelength. The figures reveal that although similar salient features appear across the four wavelengths, there is often little to no consensus per pixel. This implies a different contribution at each wavelength, highlighting the importance of a multiwavelength approach to cluster studies.

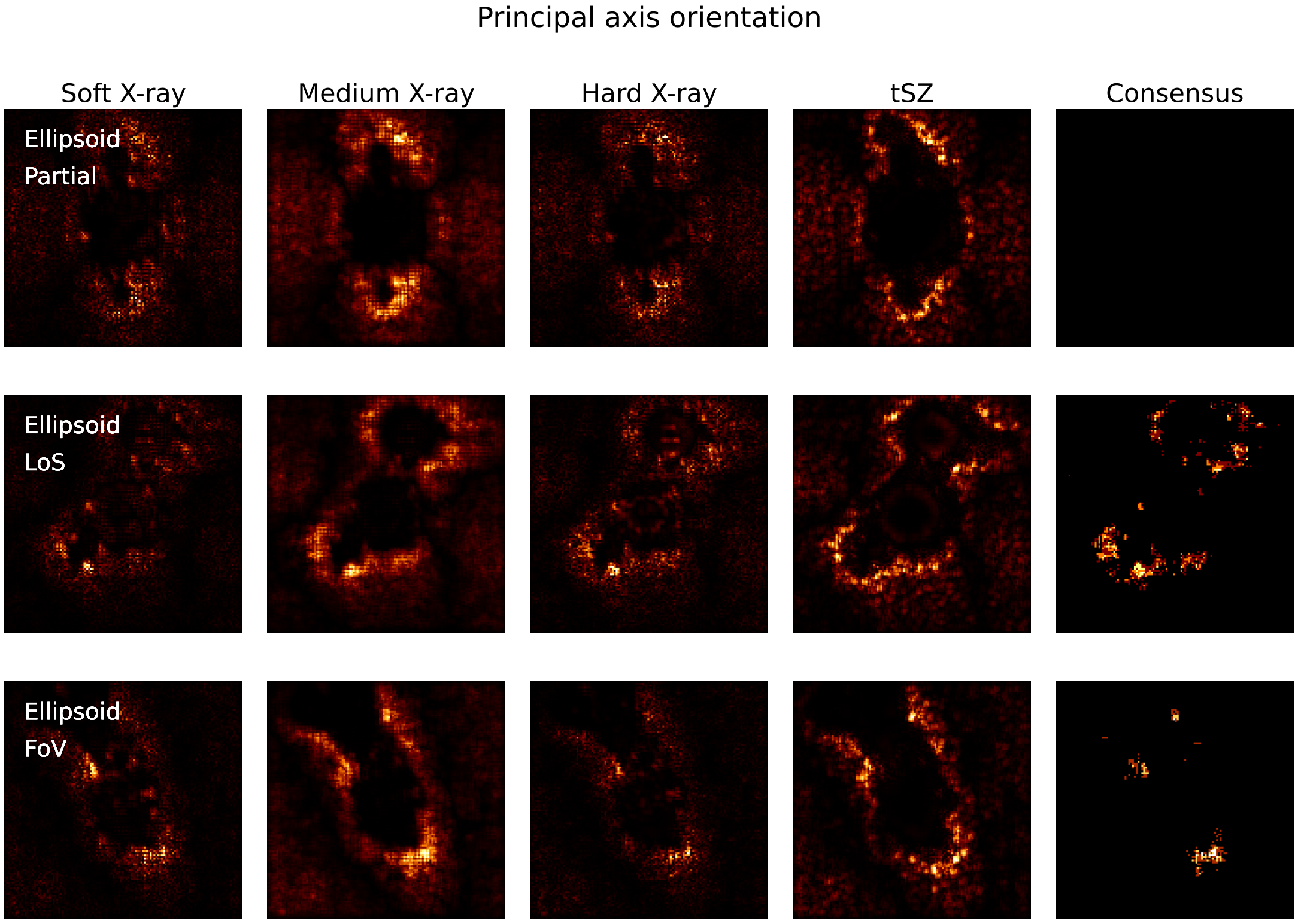

**Figure 18.** Saliency maps showing consensus activation across wavelengths for individual clusters for the prediction of principal axis orientation. Each row in this figure is a different ellipsoidal cluster whose correct principal axis orientation label is shown in the leftmost column. The rightmost column shows consensus: mean activation for pixels where there is $\geqslant$0.10 activation in every wavelength. The consensus maps reveal that although similar salient features appear across the four wavelengths, there is little to no consensus per pixel, implying that each wavelength contributes unique information.

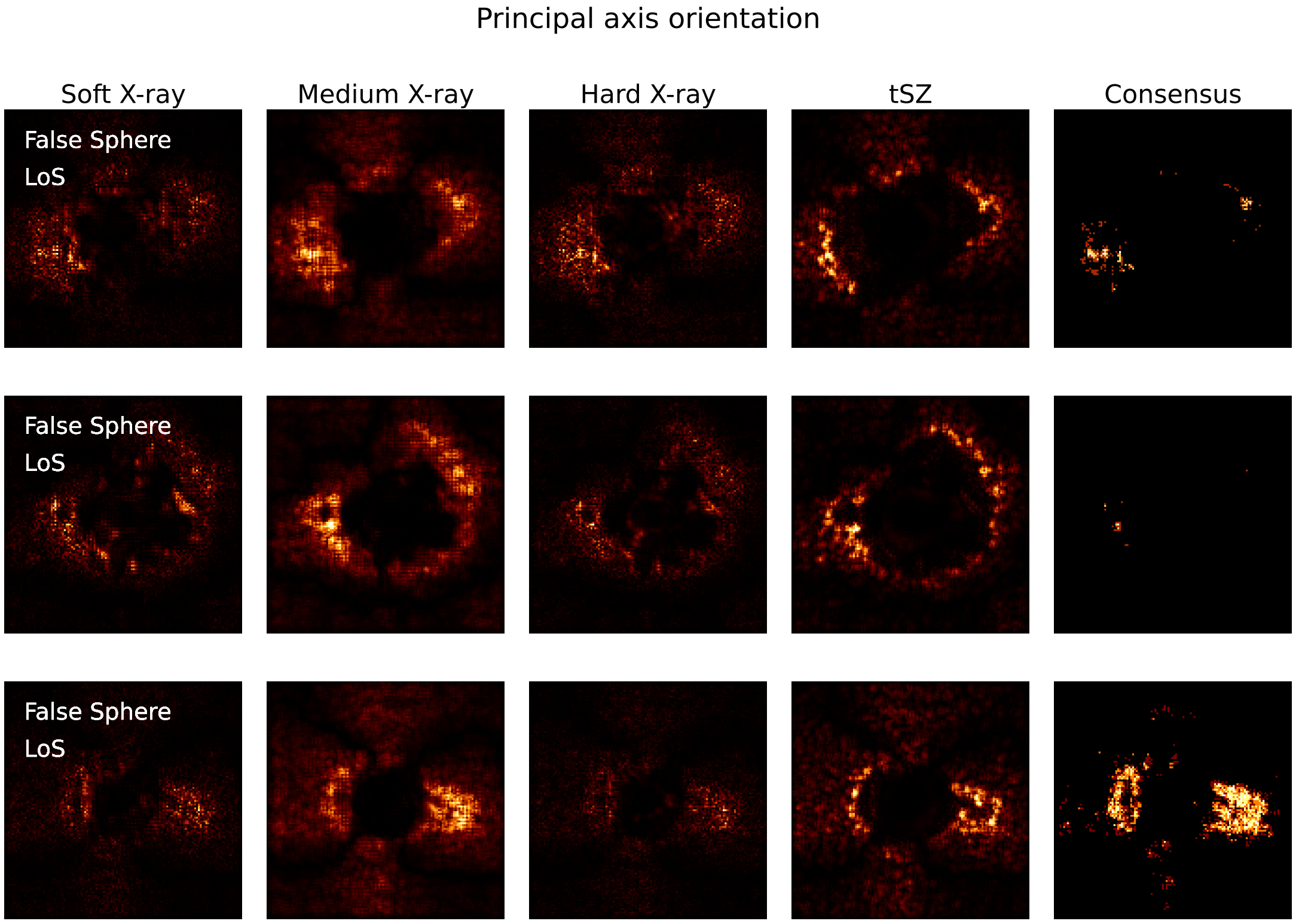

**Figure 19.** Similar to Figure 18, but for a sample of false spheres.

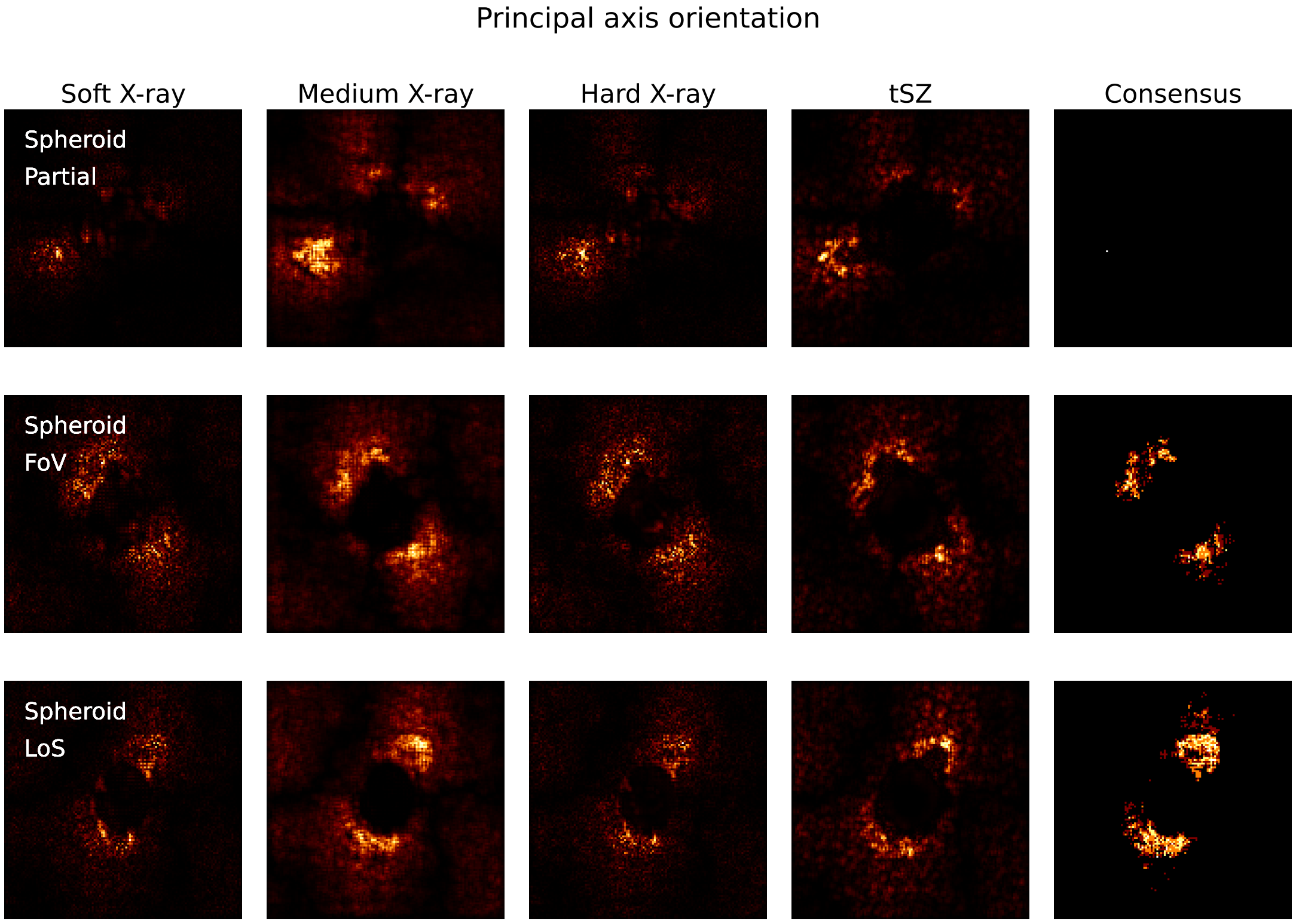

**Figure 20.** Similar to Figure 18 but for a sample of spheroids.

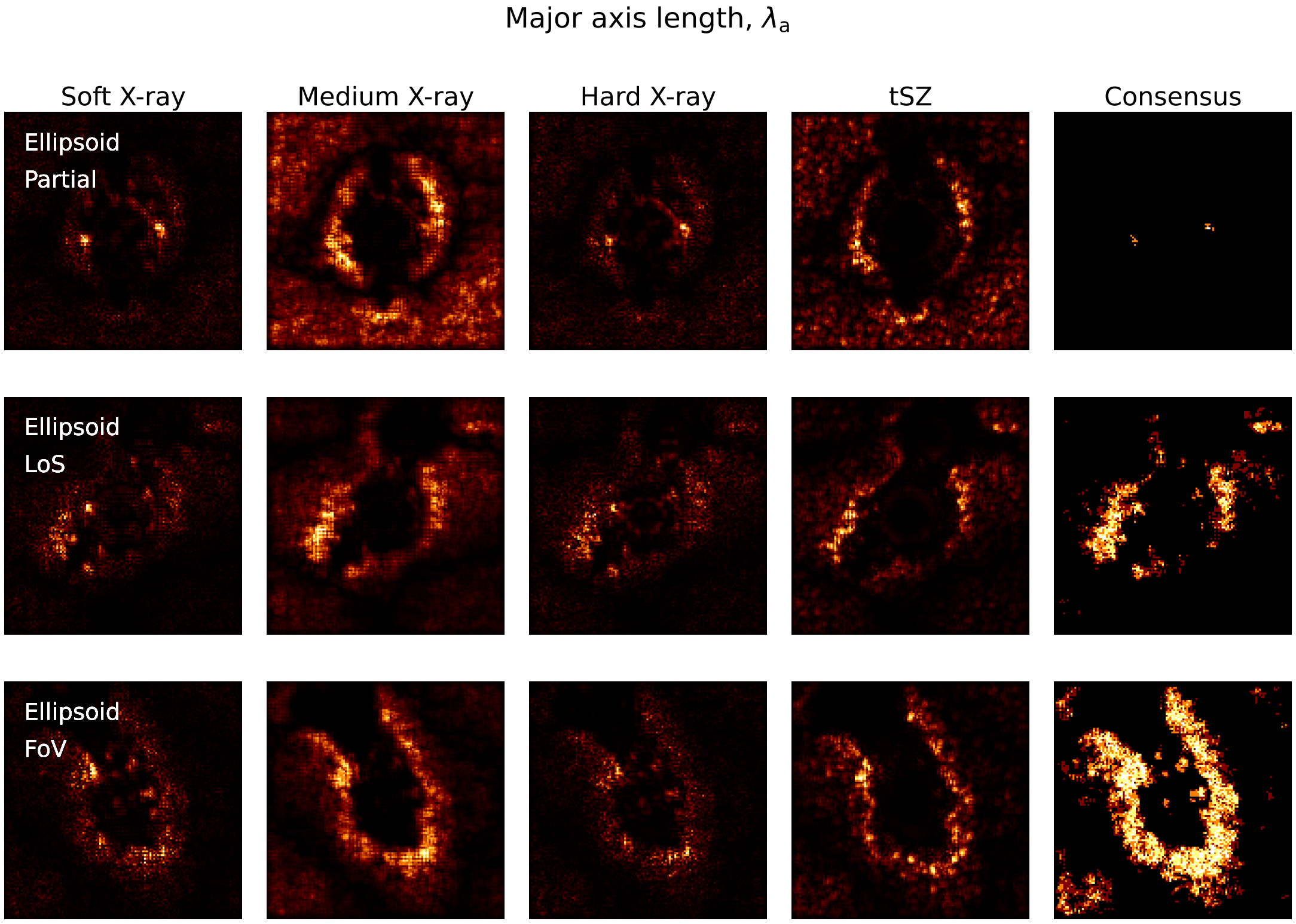

**Figure 21.** Similar to Figure 18 but for the prediction of the major axis length, $\lambda_a$.

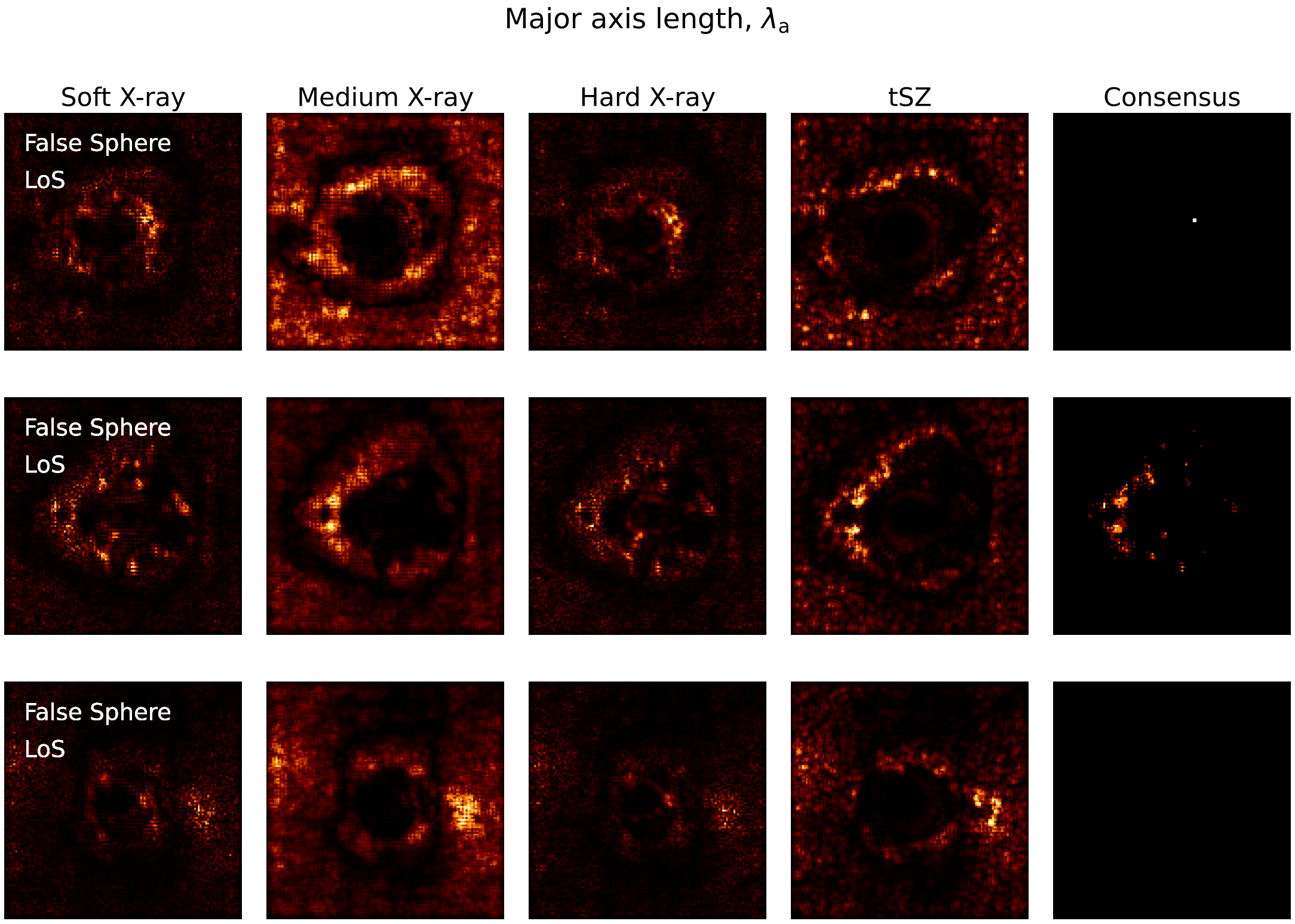

**Figure 22.** Similar to Figure 21, but for a sample of false spheres.

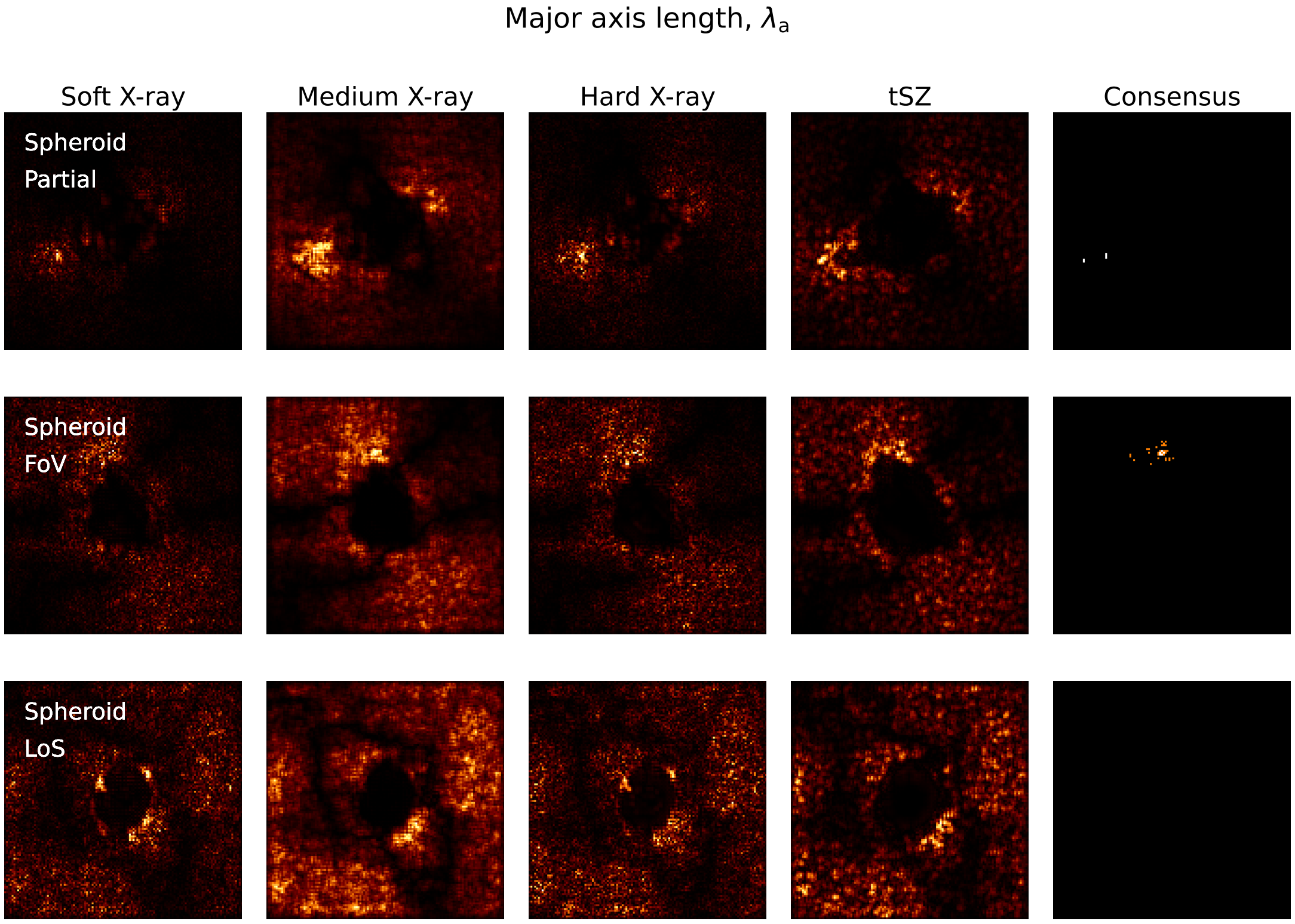

**Figure 23.** Similar to Figure 21, but for a sample of spheroids.

## ORCID iDs

Ana Maria Delgado https://orcid.org/0009-0000-2927-2104
Michelle Ntampaka https://orcid.org/0000-0002-0144-387X
Sownak Bose https://orcid.org/0000-0002-0974-5266
Lars Hernquist https://orcid.org/0000-0001-6950-1629
John Soltis https://orcid.org/0000-0002-0104-3593
John F. Wu https://orcid.org/0000-0002-5077-881X
John ZuHone https://orcid.org/0000-0003-3175-2347